\documentstyle[12pt,worldsci,epsf]{article}

\def\PL #1 #2 #3 {Phys. Lett.~{\bf#1} (#2) #3}
\def\NP #1 #2 #3 {Nucl. Phys.~{\bf#1} (#2) #3}
\def\ZP #1 #2 #3 {Z.~Phys.~{\bf#1} (#2) #3}
\def\PR #1 #2 #3 {Phys. Rev.~{\bf#1} (#2) #3}
\def\PRD #1 #2 #3 {Phys. Rev.~D {\bf#1} (#2) #3}
\def\PP #1 #2 #3 {Phys. Rep.~{\bf#1} (#2) #3}
\def\PRL #1 #2 #3 {Phys. Rev.~Lett.~{\bf#1} (#2) #3}

\def\bq{\begin{equation}}
\def\eq{\end{equation}}
\def\ba{\begin{eqnarray}}
\def\ea{\end{eqnarray}}
\def\Bsl{\hbox{/\kern-.6700em$B$}} 
\def\Dsl{\hbox{/\kern-.6700em$D$}} 
\def\Wsl{\hbox{/\kern-.6700em$W$}} 
\def\open1{\leavevmode\hbox{\xipt1\kern-3.8pt\xiipt1}}
\def\gsim{\mathrel{\raise.3ex\hbox{$>$\kern-.75em\lower1ex\hbox{$\sim$}}}}
\def\lsim{\mathrel{\raise.3ex\hbox{$<$\kern-.75em\lower1ex\hbox{$\sim$}}}}

\def\overlay#1#2{\ifmmode \setbox 0=\hbox {$#1$}\setbox 1=\hbox to\wd 0{\hss
$#2$\hss }\else \setbox 0=\hbox {#1}\setbox 1=\hbox to\wd 0{\hss #2\hss }\fi
#1\hskip -\wd 0\box 1}

\newcounter{eqletter}

\begin{document}
\baselineskip14pt
\title{
\font\fortssbx=cmssbx10 scaled \magstep2
\hbox to \hsize{
\includegraphics{uwlogo.ps}
\hskip.5in \raise.1in\hbox{\fortssbx University of Wisconsin - Madison}
\hfill$\vcenter{\hbox{\bf MADPH-95-933}
                \hbox{March 1996}}$ }\bigskip\bigskip
Developments in Perturbative QCD: Challenges from Collider Physics
\footnote{Lectures given at the {\it VIIIth J.~A.~Swieca Summer School},
Rio de Janeiro, Brazil, February 6--18, 1995.  }    }
\author{\small Dieter~Zeppenfeld\\
{\it Department of Physics, University of Wisconsin, 1150 University Ave.\\
Madison, WI 53706, USA}\\
E-mail: dieter@phenom.physics.wisc.edu}
\maketitle

\begin{center} ABSTRACT\\ [.1in]
\parbox{13.5cm}{\small
The search for new phenomena at hadron colliders requires a good understanding
of QCD processes. The analysis of multi-jet signatures in the top quark search 
at the Tevatron is one example, forward jet-tagging and rapidity gap techniques
in the analysis of weak boson scattering events at the LHC will be another 
important application. These topics are discussed in the context of 
multi-parton/multi-jet QCD processes. Also described are some
of the calculational tools, like amplitude techniques and automatic code 
generation for tree level processes.
}
\end{center}

\section{Introduction}
Ever since the start-up of the S$\bar{\rm p}$pS at CERN, hadron colliders 
have represented the high energy frontier in particle physics. Up to this day
only they have been able to produce $W$-bosons at a meaningful rate, even
though this situation will change in the summer of 1996 with the running of LEP
at an energy of 161~GeV. Top quark physics is another area which will only be
accessible to hadron colliders in the forseeable future, namely to $p\bar p$
collisions at the Tevatron for the next decade and to $pp$ collisions at
the LHC afterwards. Even the construction of an $e^+e^-$ linear collider or a
$\mu^+\mu^-$ collider will leave the LHC as the machine with the 
highest usable center of mass energy and, hence, at the forefront of discovery
physics.

The discovery of $W$ and $Z$ bosons at the S$\bar{\rm p}$pS~\cite{WZ-UA12}
with their clean leptonic decays may well represent the exception rather  
than the rule for the observation of new physics at hadron colliders. The 
top quark search~\cite{topCDF,topCDFprl,topD0},
in particular in the $t\bar t \to bW^+ \bar b W^- \to \ell^{\pm}\nu + 4$~jets
decay mode, is more representative of the complex signatures which must be
searched for in order to discover new particles. The production and decay 
of squarks and gluinos, for example, would lead to 
multi-particle final states which would give rise to  multiple jets and 
missing transverse momentum signatures or to final states with several 
leptons and jets in the case of cascade decays of the supersymmetric 
particles~\cite{tata}. 

In order to distinguish these new physics signals from Standard Model (SM) 
backgrounds, typically produced by the emission of extra, hard quarks or 
gluons in simple electroweak processes, the precise features of these 
multi-parton production processes must be determined, both for the signals 
and for the backgrounds. Fortunately, amplitude techniques have been 
developed over the last 15 years~\cite{calkul,zxu,HZ} which are very well 
suited for this task and which allow to perform parton level Monte Carlo 
studies with full tree level matrix elements. 
These lectures describe some recent uses and some of the 
developments in this field of QCD multi-parton production processes.

In Section~\ref{sec:top} the top quark search at the Tevatron will be 
discussed as an example, showing why the study of multi-parton processes 
is needed: in order to establish the signal and to measure the top-quark 
mass kinematically, the $W+4$~jets background, with and without b-quarks 
in the final state, had to be understood. Motivated by this example, 
Section~\ref{sec:MC} then provides an 
introduction into the calculation of multi-jet cross sections. I start
with a basic discussion of parton level Monte Carlo programs and then 
describe a particular amplitude technique, developed by Hagiwara and 
myself~\cite{HZ}, for the efficient evaluation of partonic cross sections. 
This technique has been encoded in the HELAS program package~\cite{HELAS} 
and is being used by MadGraph~\cite{MadGraph} to
automatically generate the FORTRAN code for (almost) arbitrary SM transition
probabilities. The basic ideas behind these programs will be described. 

An important application of multi-parton processes is the study of weak boson
scattering at the LHC~\cite{bagger}, which includes Higgs production by weak 
boson fusion, e.g. the process  $qq\to qqH\to qqW^+W^-$. In 
Section~\ref{sec:qqtoqqVV} the 
production and subsequent decay $H\to W^+W^-$ of a heavy Higgs boson will 
be used as an example to discuss some of the techniques, like forward 
jet-tagging\cite{Cahn,Froid,BCHP,BCHZ} or central jet 
vetoing\cite{BCHP,BCHZ}, 
which have been developed to distinguish weak boson scattering 
signals from QCD background processes. 

Of particular interest here are the different color structures of weak 
boson scattering (no color exchange in the $t$-channel) as compared to 
typical backgrounds where color is exchanged between the two incident partons.
This different color structure leads to very different patterns of radiated 
gluons\cite{troyan,bjgap}. 
In the signal, gluons are emitted at rather low transverse momenta 
and mainly in the forward and backward directions, leaving a central region 
with little hadronic activity apart from the Higgs decay products. Typical 
background processes emit gluons at considerably higher transverse momenta 
and preferentially in the central region. These differences suggest 
techniques like a minijet veto\cite{bpz} or rapidity gaps\cite{bjgap} 
to enhance the signal versus the backgrounds. They will be studied in 
Section~\ref{sec:qqtoqqVV}. It is helpful, however, to first consider
gluon radiation patterns and the consequences of $t$-channel color singlet
exchange in a simpler environment, namely dijet events at the Tevatron.
These topics and the recent observation of rapidity gaps at the 
Tevatron\cite{D0gap1,CDFgap,D0gap2} will be discussed in 
Section~\ref{sec:rapgap}.

\section{Jets in the Top Quark Search \label{sec:top}}

The discovery of the top quark\cite{topCDFprl,topD0} 
at the Fermilab Tevatron provides a beautiful example for the use of hadronic 
jets as a tool for discovering new particles. Let us have a brief look at
the top quark search at the Tevatron from this particular viewpoint: the 
use of multi-parton cross section calculations as a necessary ingredient 
in particle searches.

In $p\bar p$ collisions at the Tevatron, at a center of mass energy of 1.8~TeV,
the top quark is produced via quark anti-quark annihilation, $q\bar q \to
t\bar t$, and, less importantly, via $gg\to t\bar t$. Production cross
sections have been calculated at next-to-leading order\cite{ttNLO} and are
expected to be around 4~pb for a top mass of 180~GeV\cite{topcross}.
The large top decay width which is expected in the SM,
\bq
\Gamma (t\to W^+ b) \approx 1.7\;{\rm GeV}\;,
\eq
implies that the $t$ and $\bar t$ decay well before hadronization, and the 
same is true for the subsequent decay of the $W$ bosons. Thus, a parton 
level simulation for the complete decay chain, including final parton 
correlations, is a reliable means of predicting detailed properties of the 
signal. In order to distinguish the top quark signal, 
$t\bar t\to bW^+\;\bar bW^-$, from multi-jet backgrounds, 
the leptonic decay $W\to\ell\nu$ ($\ell=e,\;\mu$) of at least one of the two 
final state $W$s is required. Since the leptonic decay of both $W$s has a 
branching ratio of $\approx 4\%$ only, the prime top search channel is the 
decay chain
\bq
t\bar t\;\to\; bW^+\;\bar bW^-\;\to\; \ell^\pm\nu\; q\bar q\; b\bar b\; ,
\label{eq:topsig}
\eq
which, within the SM, has an expected branching ratio of $\approx 30\%$. After 
hadronization each of the final state quarks in (\ref{eq:topsig}) may 
emerge as a hadronic jet, provided it carries enough energy. Thus the 
$t\bar t$ signal is expected in $W+3$~jet and $W+4$~jet 
events\footnote{Gluon bremsstrahlung may increase the number of jets further 
and thus all $W+\geq 3$~jet events are potential $t\bar t$ candidates.}.

\begin{figure}[htb]
\epsfxsize=4.0in
\epsfysize=4.0in
\begin{center}
\hspace*{0in}
\epsffile{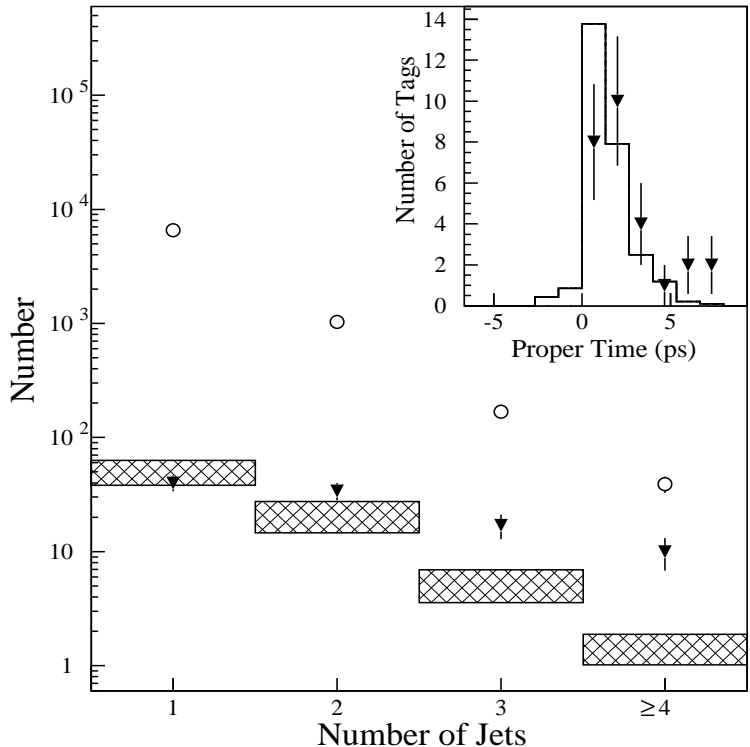}
\vspace*{-0.1in}
\caption{
Number of $W+n$~jet events in the CDF top quark search as a function of jet 
multiplicity. Number of observed events are given without $b$-tagging 
(open circles) and with an SVX tag (triangles). The expected background, 
mainly from QCD $W+n$~jet events, is given by the cross-hatched bars.
{}From Ref.~3. 
\label{fig:cdftop1}
}
\vspace*{-0.1in}
\end{center}
\end{figure}

Events with leptonic $W$ decays and several jets can also arise from QCD 
corrections to the basic Drell-Yan process $q\bar q\to W^\pm \to \ell^\pm\nu$.
The process $ug\to dggW^+$, for example, will give rise to $W+3$~jet events
and its cross section and the cross sections for all other subprocesses with 
a $W$ and three partons in the final state need to be calculated in order 
to assess the QCD background of $W+3$~jet events, at tree level. $W+n$~jet 
cross sections have been calculated for $n=3$ jets\cite{W3j} and $n=4$ 
jets\cite{W4j} and some of the methods used in these calculations will 
be discussed in Section~\ref{sec:MC}. As in the experiment, the calculated
$W+n$~jet cross sections depend critically on the minimal transverse energy
of a jet. CDF, for example, requires a cluster of hadrons to carry 
$E_T>15$~GeV to be identified as a jet\cite{topCDF}, and this observed 
$E_T$ must then be translated into the corresponding parton transverse 
momentum in order to get a prediction for the $W+n$~jet cross sections. 

At this level the QCD backgrounds are still too large to give a viable top
quark signal. The situation is improved substantially by using the fact that 
two of the four final state partons in the signal are $b$-quarks, while only 
a small fraction of the $W+n$~parton background events have $b$-quarks in the 
final state. These fractions are readily calculated by using $W+n$~jet Monte 
Carlo programs. There are several experimental techniques
to identify $b$-quark jets, all based on the weak decays of the produced
$b$'s. One method is to use the finite $b$ lifetime of about $\tau=1.5$~ps 
which leads to $b$-decay vertices which are displaced by 
$\gamma c\tau=$~few mm from the primary interaction vertex. These 
displaced vertices can be resolved by precision tracking, with the aid of 
their Silicon VerteX detector in the case of CDF, and the method is, 
therefore, called  SVX tag. In a second method, $b$ decays are identified 
by the soft leptons which arise in the weak decay chain 
$b\to W^*c,\; c\to W^*s$, where either one of the virtual $W$s may decay 
leptonically\cite{topCDF,topD0}.

The combined results of using jet multiplicities and SVX $b$-tagging to
isolate the top quark signal are shown in Fig.~\ref{fig:cdftop1}. A clear
excess of $b$-tagged 3 and 4 jet events is observed above the expected 
background. The excess events would become insignificant if all jet 
multiplicities were combined or if no $b$-tag were used (see open circles).
Thus jet counting and the identification of $b$-quark jets have been 
critical for the discovery of the top quark.

\begin{figure}[htb]
\epsfxsize=4.0in
\epsfysize=4.0in
\begin{center}
\hspace*{0in}
\epsffile{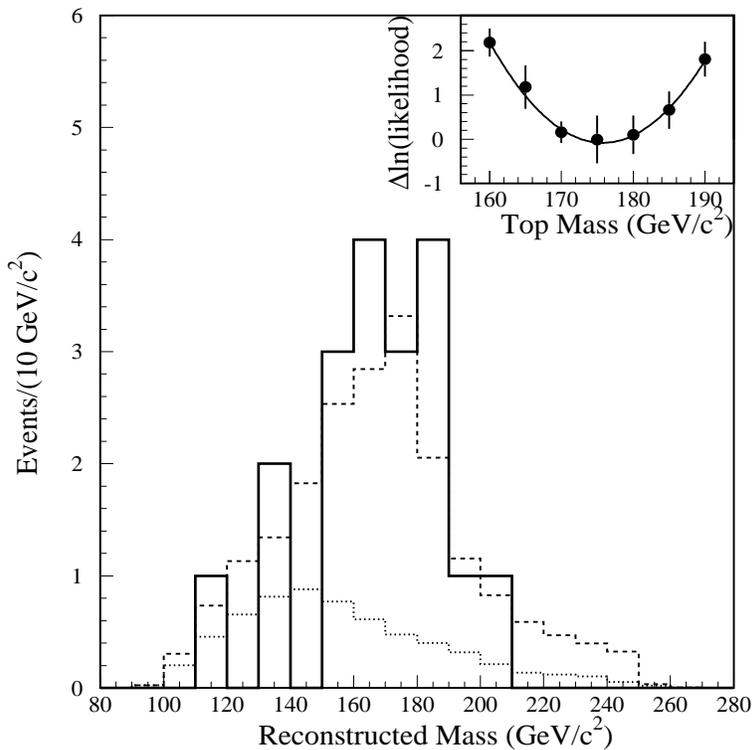}
\vspace*{-0.1in}
\caption{
Reconstructed  mass distribution for $W+\geq 4$~jet events with a
$b$-tag. The solid histogram represents the CDF data. Also included are the
expected background (dotted histogram) and the expected signal+background
for $m_t = 175$~GeV. The insert shows the likelihood fit to determine the top 
quark mass, which yielded $m_t=176\pm 8\pm 10$~GeV. From 
Ref.~3. 
\label{fig:cdfmt}
}
\vspace*{-0.1in}
\end{center}
\end{figure}

Beyond counting the number of jets above a certain transverse energy, the
more detailed kinematic distributions, their summed scalar $E_T$'s\cite{topD0}
and multi-jet invariant masses, have also been critical in the top quark 
search and thus needed to be predicted for the $W+n$~jet backgrounds. 
The top quark mass determination, for example, relies on a good understanding
of these distributions. Ideally, in a $t\bar t\to (\ell^+\nu b)(q\bar q\bar b)$
event, for example, the two subsystem invariant masses should be equal to the
top quark mass,
\bq 
m_t \approx m(\ell^+\nu b) \approx m(q\bar q\bar b)\; .
\eq
Including measurement errors, wrong assignment of observed jets to the two
clusters, etc. one needs to perform a constrained fit to extract $m_t$. The
CDF result of this fit\cite{topCDFprl} is shown in Fig.~\ref{fig:cdfmt}. 
Combining this with the corresponding D0 measurement\cite{topD0} gives the 
value\cite{kleink}
\bq 
m_t = 180 \pm 13\; {\rm GeV}\; .
\eq
In addition, Fig.~\ref{fig:cdfmt} demonstrates that the observed $b$-tagged
$W+4$~jet events (solid histogram) are considerably harder than the QCD
background (dotted histogram). On the other hand the data agree very well 
with the top quark hypothesis (dashed histogram). The kinematic 
distributions of the jets contain much valuable information and in order 
to extract it we must be able to reliably calculate
them for processes with many partons in the final state.

\section{Monte Carlos and Amplitude Techniques\label{sec:MC}}
In the previous Section we have seen that the isolation of new physics 
signals like top quark pair production requires precise predictions for 
the properties of signal and background processes. One needs to 
know how the corresponding differential cross sections change as a function 
of the kinematic variables of the observed leptons and jets. Starting with
a generic process
\bq
p\bar p \to B_1B_2\dots B_nX\; ,
\label{eq:proc-jets}
\eq
where the $B_i$ stand for the final state leptons and jets, we will discuss 
in the following how to determine the corresponding cross sections in a 
parton level Monte Carlo program, at tree level. We begin by considering 
the general structure of a Monte Carlo program in Section~3.1. A formalism 
for the fast numerical evaluation of polarization amplitudes is described in
Section~3.2, and Section~3.3 discusses automatic code generation for the
calculation of squared matrix elements, as implemented in the 
Madgraph\cite{MadGraph} program.

\subsection{Monte Carlo simulations at the parton level}
At the parton level, the process $p\bar p \to B_1B_2\dots B_nX$ 
originates from a number of different subprocesses,
\bq
a_1a_2 \to b_1b_2\dots b_nX\; ,
\label{eq:proc-partons}
\eq
and the cross sections for all these subprocesses must be added to obtain
the probability for observing the final state $B_1B_2\dots B_n$. 

Consider the $B_1B_2=W^+j$ signature as an example. The contributing parton 
level subprocesses $a_1a_2 \to b_1b_2$ are
\ba
\bar d u,\; \bar s c,\; u\bar d,\; c\bar s & \to & W^+ g\; , 
\label{eq:Wj-1}\\
u g,\; g u                                 & \to & W^+ d\; , 
\label{eq:Wj-2}\\
c g,\; g c                                 & \to & W^+ s\; , 
\label{eq:Wj-3}\\
\bar d g,\; g \bar d                       & \to & W^+ \bar u\; ,  
\label{eq:Wj-4}\\
\bar s g,\; g \bar s                       & \to & W^+ \bar c\; ,  
\label{eq:Wj-5}
\ea
where Cabibbo mixing has been neglected for simplicity\footnote{Because of the 
small value of the Cabibbo angle, sin$^2\theta_C=0.049$, the resulting error
is negligible compared to the neglect of higher order QCD corrections, for 
example. It is further mitigated by the fact that after summing over
final state flavors and neglecting quark masses, only the annihilation 
processes (\ref{eq:Wj-1}) depend on the value of the Cabibbo angle.}. 
The various subprocesses are closely related, of course: by choosing either 
the proton or the anti-proton as the source of the initial state quark or 
anti-quark, by interchanging first and second generation quark flavors, or 
by crossing the gluon between the initial and final states. Using these
relations, the required 
analytical calculations can be reduced significantly. At the numerical level,
however,  only the flavor independence of QCD corrections allows to reduce 
the number of independent cross section evaluations.

In order to calculate the cross section for the $B_1B_2\dots B_n$ final state,
the squared matrix elements, summed over final state colors and polarizations
and averaged over initial ones, 
\bq
\overline{\sum}|{\cal M}|^2 = {1\over 4}\;{1\over \# {\rm colors}(a_1a_2)}\;
\sum_{\rm colors}\;\;\sum_{\rm polarizations} |{\cal M}|^2
\label{eq:sumbarM2}
\eq
must be known for all contributing subprocesses. Their determination is the 
subject of the next two subsections. The full cross section is then given by
\ba
\sigma & = & \int dx_1dx_2 \sum_{\rm subprocesses}\; 
             f_{a_1/p}(x_1)\; f_{a_2/\bar p}(x_2)\; \nonumber \\  
&& {1\over 2\hat s}\int d\Phi_n(x_1p+x_2\bar p;\;p_1\dots p_n) 
   \Theta({\rm cuts}) 
\;\;\; \overline{\sum}|{\cal M}|^2(a_1a_2\to b_1b_2\dots b_n)\; .
\label{eq:sigma}
\ea
Here $f_{a_1/p}(x_1)$ is the probability to find parton $a_1$ inside the 
proton, carrying a fraction $x_1$ of the proton momentum, i.e. the $a_1$ 
momentum is $x_1p$. Similarly, $f_{a_2/\bar p}(x_2)$ is the parton $a_2$ 
distribution function inside the anti-proton. In the second line of
(\ref{eq:sigma}) $1/2\hat s$ is the flux factor for the partonic cross 
section, and
\bq
d\Phi_n(P;\;p_1\dots p_n) = 
\prod_{i=1}^n \left( {d^3{\bf p}_i\over (2\pi)^3 2E_i} \right)
(2\pi)^4\;\delta^4(P-\sum_i p_i)
\eq
is the Lorentz invariant phase space element. Finally, $\Theta({\rm cuts})$ 
is the acceptance function, which summarizes the kinematical cuts on all the 
final state particles, i.e. $\Theta = 1$ if all the partons $a_i$ satisfy all 
the acceptance cuts and $\Theta=0$ otherwise.

The calculation of the cross section in eq.~(\ref{eq:sigma}) involves a 
$3n-2$ dimensional integral which is best performed by Monte Carlo. In 
order to do this one first maps the integration region onto a $3n-2$ 
dimensional hypercube of unit length, i.e. one rewrites the integration 
measure as
\bq
{1\over 2\hat s} dx_1 dx_2 d\Phi_n = J\prod_{i=1}^{3n-2} dr_i\; ,
\eq
where $J$ is the Jacobian factor of the transformation of integration 
variables. The Monte Carlo integration then samples the integrand randomly
at some large number, $N$, of points in the hypercube. In other words, 
take $N$ sets $\{ r_i\}$ of $3n-2$ random numbers each. A good approximation 
to the cross section is then provided by
\bq
\sigma \approx {1\over N} \sum_{\{ r_i\}}\; J\; \sum_{subprocesses}\;
f(x_1)f(x_2) \overline{\sum} |{\cal M}|^2\;\Theta({\rm cuts})\;,
\label{eq:MC}
\eq
which approaches the correct cross section value in the limit $N\to\infty$.

Eq.~(\ref{eq:MC}) contains all the elements of a Monte Carlo program:
\begin{itemize}
\item{}
An integration routine (for example the programs VEGAS\cite{VEGAS} or 
BASES\cite{BASES}) provides the set $\{ r_i\}$ of random numbers and the 
weight factor $wgt=``1/N''$. Each set $\{ r_i\}$ is called an event.
\item{}
The phase space generator generates the kinematic variables, i.e. the parton 
momenta $p_i$, the initial parton momentum fractions $x_1$ and $x_2$, 
and the Jacobian factor $J$ from this set of random numbers.
\end{itemize}

\noindent
As a trivial example consider the two body phase space in the decay process
$a(P)\to b_1(p_1)b_2(p_2)$, where  $b_1$ and $b_2$ represent two massless 
particles. The Lorentz invariant phase space element is best 
written in polar coordinates in the $P$-rest frame:
\bq
d\Phi_2(P;p_1,p_2) = {1\over 32\pi^2}\;d\Omega={dr_1\;dr_2\over 8\pi}\;,
\eq
where the components of the 4-momentum $p_1$ are given by the two random 
numbers $r_1$ and $r_2$ via
\ba
p_1&=&{\sqrt{P^2}\over 2}(1,\;{\rm sin}\;\theta\;{\rm cos}\;\phi,\;
{\rm sin}\;\theta\;{\rm sin}\;\phi,\;{\rm cos}\;\theta)\;, \\
{\rm cos}\;\theta &=& 2r_1-1\;,\;\; \phi=2\pi r_2\;.
\ea
Since the phase space element $d\Phi_2$ is Lorentz invariant, the momenta 
may now be boosted to an arbitrary frame and the momentum $p_2$ is then 
calculated from momentum conservation,
\bq
p_2=P-p_1\;.
\eq
\begin{itemize}
\item{}
Given the 4-momenta of the final state particles $\{p_i\}$ one can easily 
check if the event passes the acceptance cuts, i.e. $\Theta({\rm cuts})=1$. 
If the event is accepted then the squared matrix elements, 
$\overline{\sum} |{\cal M}|^2$, are calculated for all contributing 
subprocesses.
\item{}
The weight factor (value of the integrand)
\bq
w = J\; \sum_{subprocesses}\;
f(x_1)f(x_2) \overline{\sum} |{\cal M}|^2\; \Theta({\rm cuts})
\eq
is returned to the integration routine, which then adds these weights for 
the various regions of the hypercube, thus obtaining an estimate of the 
cross section, $\sigma$, according to eq.~(\ref{eq:MC}).

\item{}
At the same time one can generate arbitrary distributions $d\sigma/dz$ 
by filling a number of histograms. For each event one uses the 4-momenta 
$p_i$ to determine the value of $z$ for each histogram in question. Given 
the bin width $\Delta z$ one adds $w\cdot wgt/\Delta z$ to the bin 
corresponding to $z$ and thus obtains a correctly normalized histogram 
of $d\sigma/dz$ at the end. 
\end{itemize}

In the above procedure the choice of mapping from the unit hypercube to the 
Lorentz invariant phase space element is obviously ambiguous. The art of 
writing a good phase space generator is to choose a mapping which results in
a large fraction of events having a sizable weight, close to the maximum 
weight $w_{max}$ encountered in the full sample. Events with small weights 
contribute very little to $\sigma$ in eq.~(\ref{eq:MC}) and thus do not 
reduce the statistical error, but they require as much time to be computed
as high weight events. The task in writing the phase space generator is thus
to anticipate the dominant phase space regions, those regions in which poles
in the matrix elements lead to large cross sections, and to find a mapping
which smooths out these maxima in the product 
$J\;\overline{\sum} |{\cal M}|^2$. It is a good practice to histogram the 
distribution of the weight factors, $w$, in order to judge whether 
improvements to the phase space generator should be made. 

\subsection{Amplitude techniques}
It is the squared amplitudes, $\overline{\sum} |{\cal M}|^2$, which contain 
the full information on the dynamics of the underlying physical processes and 
which are at the core of any cross section calculation. For each subprocess
$a_1(k_1)\;a_2(k_2)\to b_1(p_1)\dots b_n(p_n)$ and phase space point 
$(k_1,k_2;p_1,\dots,p_n)$ the squared amplitude has a well defined 
numerical value. The question then is: What is the most efficient way to 
``evaluate the Feynman graphs'' and to thus obtain this value?

Traditionally, the method of choice has been to use ``trace techniques'' to
express $\overline{\sum} |{\cal M}|^2$ in terms of scalar products 
$p_i\cdot p_j$. For relatively simple processes, like $2\to 2$ scattering,
one thus obtains very compact and easy to interpret analytical expressions,
in particular for unpolarized scattering, and no residual polarization and 
color sums need to be performed in evaluating Eq.~(\ref{eq:sumbarM2}) 
numerically. Unfortunately, the expressions become very large as the 
number, $f$, of individual Feynman graphs increases. For
\bq
{\cal M} = \sum_{i=1}^f {\cal M}_i
\eq
a total of $f(f+1)/2$ cross terms Re(${\cal M}_i^*{\cal M}_j$) need to be 
calculated and hence the complexity of the problem grows at least with the 
square of the number of Feynman graphs. For $2\to 4$ or $2\to 5$ processes 
the task typically becomes forbidding.

For such complex processes the resulting analytical expressions cannot easily 
be interpreted any more, and in practice one merely needs them for subsequent 
numerical evaluation. It then becomes advantageous to numerically 
evaluate the individual Feynman amplitudes ${\cal M}_i$ as $f$ complex
numbers which can be summed and squared trivially. The complexity of this 
problem only grows linearly with $f$. 

A number of approaches have been developed for the direct evaluation of 
polarization amplitudes. The CALKUL method\cite{calkul} and improvements on 
it\cite{zxu} may be the most widely used and it yields rather compact 
expressions for helicity amplitudes involving massless fermions. Similar
(high) numerical speeds are achieved via the HZ method\cite{HZ} which was 
developed with massive fermions in mind, but which simplifies significantly 
as well when massless fermions are considered. As compared to CALKUL results,
helicity amplitudes in the HZ method are typically not as compact. However,
the basic structure of the underlying Feynman graphs is preserved and this 
fact has proven invaluable in the automation of code 
generation\cite{MadGraph}. At the same time it becomes very easy to add new 
features to existing calculations, like implementing finite width 
effects\cite{BZ} or anomalous gauge boson couplings\cite{HPZH}. 

The HZ method is implemented in a FORTRAN77 package called
HELAS\cite{HELAS}. In the following the principles behind the method will 
be presented. For full details the reader is referred to the
original articles\cite{HZ,HELAS}.

In order to exploit the simplifications which occur for massless fermions
we work in the chiral representation, with $\gamma_5$ given by
\bq
\gamma_5 = \left(\begin{array}{rc}
-\open1 & 0 \\
0 & \open1 
\end{array}\right) \,.
\eq 
External spinors are split into their chiral components,
\bq
\psi = \left(\begin{array}{c}
\psi_- \\ 
\psi_+
\end{array}\right) \,,
\eq
where $\psi_-$ and $\psi_+$ are two-component Weyl spinors of negative and 
positive helicity, respectively. As an example consider the wave function 
of an incoming fermion of 4-momentum 
$p^\mu=(p^0,p_x,p_y,p_z)=(p^0,{\bf p})$ and helicity $\sigma/2$. It can be 
represented by a $u$-spinor with chiral components
\bq
u(p,\sigma)_\pm = \sqrt{p^0\pm\sigma |{\bf p}|}\;\chi_\sigma(p) \;.
\label{eq:u-spinor}
\eq
Here the Weyl spinors of fixed helicity $\sigma/2 = \pm 1/2$ are given
explicitly by
\ba
\chi_+(p) & = & {1\over \sqrt{ 2|{\bf p}|(|{\bf p}|+p_z)}}
\left(\begin{array}{c}
|{\bf p}| + p_z \\
p_x + ip_y
\end{array}\right) \,, \\
\chi_-(p) & = & {1\over \sqrt{ 2|{\bf p}|(|{\bf p}|+p_z)}}
\left(\begin{array}{r}
-p_x + ip_y \\
|{\bf p}| + p_z
\end{array}\right) \,.
\ea
Given this explicit representation, the four components of the $u$-spinor can
be calculated as a set of four complex numbers once the external 4-momentum
and the helicity are fixed. In the HELAS package this is achieved by a simple
subroutine call,

{\tt call IXXXXX(P,FMASS,NH,+1,PSI)}

\noindent
where {\tt P=P(0:3)} is the external fermion 4-momentum, {\tt FMASS} is the
fermion mass, {\tt NH}$=\sigma=\pm 1$ its helicity, the entry +1 indicates 
that an external fermion
(as opposed to anti-fermion) wave function needs to be calculated and the 
output {\tt PSI = PSI(1:6)} contains the resulting spinor $\psi=u(p,\sigma)$ 
and the 4-momentum $p$ stored as 
\bq
{\tt PSI} = (\psi_1,\;\psi_2,\;\psi_3,\;\psi_4,\;
p^0+ip_z,\;p_x+ip_y)\;. 
\eq

\noindent
In the same way the wave function $\overline{v}(p,\sigma)$ of an 
anti-fermion in the initial state is calculated via

{\tt call OXXXXX(P,FMASS,NH=$\sigma$,$-1$,PSI)}

\noindent
where the entry $-1$ indicates that the wave function of an anti-fermion is 
needed and the name {\tt OXXXXX} refers to the direction of the arrow on the
fermion line in the Feynman graph: an outgoing fermion line is considered. 

The other wave functions which are needed in the calculation of Feynman 
amplitudes are the polarization vectors of external vector bosons. Given the
4-momentum $k^\mu = (E,\;k_x,\;k_y,\;k_z)=(E,{\bf k})$ of a vector boson
of mass $m$, three independent polarization vectors which satisfy 
$k\cdot\varepsilon$=0 are given by ($k_T = \sqrt{k_x^2+k_y^2}$)
\ba
\varepsilon^\mu(k,0)&=& 
{E\over m|{\bf k}|}({\bf k}^2/E,\;k_x,\;k_y,\;k_z)\;,\\
\varepsilon^\mu(k,1)&=& 
{1\over |{\bf k}|k_T} (0,\;k_xk_z,\;k_yk_z,\;-k_T^2)\;,\\
\varepsilon^\mu(k,2)&=& {1\over k_T}  (0,\;-k_y,\;k_x,\;0)\;.
\ea
The polarization vectors for fixed helicities $\lambda=\pm 1$ are obtained 
from this Cartesian basis via
\bq
\varepsilon^\mu(k,\lambda=\pm 1) =
{1\over\sqrt{2}}\left(\mp\varepsilon^\mu(k,1)-i\varepsilon^\mu(k,2)\right)
\eq
and they are obtained in HELAS via subroutine calls

{\tt call VXXXXX(K,VMASS,NHEL($=\lambda$),$-1$,EPS)}

\noindent
where $-1$ indicates that the wave function $\varepsilon^\mu$ for a vector 
boson in the initial state is to be calculated (+1 would be appropriate for
calculating the wave function $\varepsilon^{\mu *}$ of a final state vector 
boson) and as for fermions {\tt EPS=EPS(1:6)} combines the four components of
the complex polarization vector with momentum information on the vector boson.

Having determined the external particle wave functions we need to do the 
$\gamma$-matrix algebra corresponding to the fermion lines inside Feynman 
graphs. For any 4-vector $a^\mu$ the contraction with $\gamma_\mu$, in the 
chiral representation, can be written as
\bq
\hbox{/\kern-.4900em$a$} = a^\mu\gamma_\mu = 
\left(\begin{array}{cc}
0 & (\overlay a/)_+\\
(\overlay a/)_- & 0
\end{array}\right) \,.
\eq 
Here the $2\times 2$ submatrices (\hbox{/\kern-.4900em$a$})$_\pm$ are 
given by
\bq
(\hbox{/\kern-.4900em$a$})_\pm= \left(\begin{array}{cc}
a^0 \mp a_z & \mp(a_x - ia_y)\\
\mp(a_x+ia_y) & a^0 \pm a_z
\end{array}\right)
\eq

\begin{figure}[t]
\epsfxsize=3.0in
\epsfysize=1.0in
\begin{center}
\hspace*{0in}
\epsffile{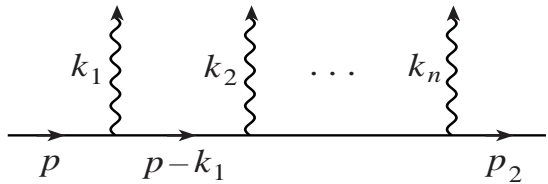}
\vspace*{-0.1in}
\caption{
Piece of a Feynman graph corresponding to vector boson emission off a 
fermion line.     \label{fig:def-ket}
}
\vspace*{-0.1in}
\end{center}
\end{figure}
Now consider external vector bosons attached to a fermion line as depicted in 
Fig.~\ref{fig:def-ket}. The external fermion spinor $\psi=\psi(p)$, the first 
vector boson emission and the fermion propagator of momentum $p-k_1$
give rise to a new four-component spinor,
\bq
|k_1,\;p> = {i\over \overlay{p\,}/-\overlay k/_1-m_f}\overlay\varepsilon/_1\;
\left( -ig_L{1-\gamma_5\over 2}-ig_R{1+\gamma_5\over 2}\right)\; \psi\;,
\eq
where the left- and right-handed couplings $g_L$ and $g_R$ depend on the 
nature of the vector boson considered. In our explicit representation of 
Dirac-matrices and spinors the resulting ket is obtained by simple 
$4\times 4$ matrix multiplication which is performed numerically in HELAS
via a subroutine call

{\tt call FVIXXX(PSI,EPS1,G,FMASS,0,PSIV)}

\noindent
Here {\tt PSI} and {\tt EPS1} contain the full information on the external
spinor and the 
vector boson of momentum $k_1$, $G = (-g_L,-g_R)$ stands for the coupling 
constants, the 0 indicates that the fermion decay width has been set to zero
and the output {\tt PSIV = PSIV(1:6)} again stores the four-component
spinor $|k_1,\;p>$ and the 4-momentum $p-k_1$ on the 
outgoing spinor line, encoded in terms of two complex numbers.

The emission of additional vector bosons now becomes trivial: the output
{\tt PSIV} of the last call of {\tt FVIXXX} is used as the input {\tt PSI} of
the next {\tt FVIXXX} call, with vector boson 1 replaced by vector boson 2, 
etc. This leaves us with the last vector boson vertex, 
\bq
\overline{\psi}(p_2){\overlay\varepsilon/}_n
\left( -ig_L{1-\gamma_5\over 2}-ig_R{1+\gamma_5\over 2}\right)
|k_1,\dots,k_{n-1},\;p>\; ,
\eq
i.e. the task of calculating the final amplitude, corresponding to the 
Feynman graph of Fig.~\ref{fig:def-ket}, from the complex 4-spinor 
$|k_1,\dots,k_{n-1},\;p>$, the complex conjugate spinor 
$\overline{\psi}(p_2)$ and a 4$\times$4 matrix sandwiched 
in between. This, of course, is a trivial task and accomplished in HELAS 
by the call of another subroutine, {\tt IOVXXX}. Alternatively, if the last 
vector boson is virtual, one may want to calculate the 4-vector
\bq
J^\mu = {i\over k_n^2-m_V^2}
\left(-g^{\mu\nu}+{k_n^\mu k_n^\nu\over m_V^2}\right)
\overline{\psi}(p_2)\gamma_\nu
\left( -ig_L{1-\gamma_5\over 2}-ig_R{1+\gamma_5\over 2}\right)
|k_1,\dots,k_{n-1},\;p>\; ,
\eq
which can then be used in place of the polarization vector for an external 
vector boson in e.g. the calculation of a second fermion line. This is done 
by the subroutine {\tt JIOXXX}.

In an analogous way non-abelian couplings of vector bosons are handled. 
{}From the polarization vectors and momenta of two external gauge bosons (or
the corresponding currents in the case of virtual gauge bosons) one may
calculate a new current ({\tt call JVVXXX}). Alternatively, one can 
calculate an amplitude from the ``wave functions'' and momenta of 
three gauge bosons, via the call of yet another subroutine, {\tt VVVXXX}.

\begin{figure}[htb]
\epsfxsize=4.5in
\epsfysize=3.0in
\begin{center}
\hspace*{0in}
\epsffile{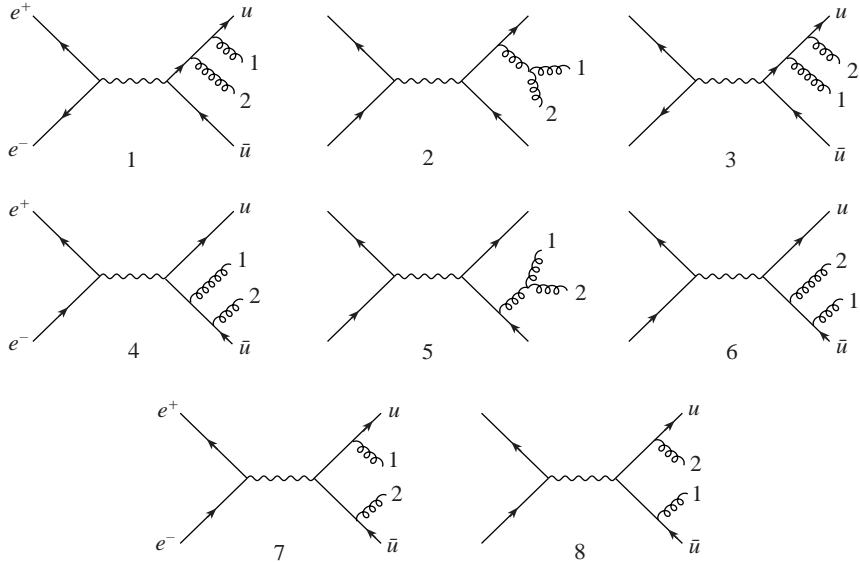}
\caption{
Feynman graphs for the process $e^+e^-\to \bar u u g g$.
\label{fig:eeuugg}
}
\vspace*{-0.1in}
\end{center}
\end{figure}

In this fashion numerical expressions for arbitrary Feynman diagrams can be
built in a block-like fashion. This method actually is quite economic, since
many of the building-blocks will be common to several Feynman graphs and
only need to be calculated once. As an example consider the Feynman graphs
for the process $e^+e^-\to \bar u u g g$ depicted in Fig.~\ref{fig:eeuugg}.
The incoming $e^+e^-$ current is common to all 8 graphs. Attaching this 
current to the external $v$-spinor of the $\bar u$ leads to a 4-spinor 
which is common to graphs 1, 2, and 3. The virtual gluon ``current'', which 
describes the 3-gluon vertex, appears in graphs 2 and 5, and so on. Exploiting 
the repetitive elements in the various Feynman graphs substantially increases
the efficiency of the numerical calculation, in particular for processes with 
a very large number of graphs.

Finally one should mention another convenient feature of the HELAS program.
When crossing initial and final state particles the fermion arrows in the
Feynman graphs do not change, only the overall signs of the 4-momenta of the 
crossed particles need to be flipped. This is achieved in HELAS by flipping
the sign factors in the determination of the external wave functions: In 
the {\tt IXXXXX, OXXXXX} and {\tt VXXXXX} calls discussed above the sign 
of the second last argument needs to be reversed. No further changes are
required when calculating crossing related processes.

\subsection{Automatic code generation: the MadGraph example}

The discussion of the previous subsection should have made it clear that 
writing the code, which evaluates the squared amplitudes for the various 
subprocesses, is actually quite a mechanical and boring exercise, exactly
the type of task which should be relegated to a computer. Indeed, a number 
of programs have been developed in recent years for automatic code 
generation for scattering processes\cite{feynarts,grace-etc}. The example to
be discussed below is MadGraph\cite{MadGraph}. MadGraph is a standard Fortran 
program\footnote{
The MadGraph\cite{MadGraph} source code and the HELAS\cite{HELAS} 
routines called by the Madgraph output can be obtained via anonymous 
ftp from {\it phenom.physics.wisc.edu} where it is located in the 
directory {\it pub/madgraph/}.} 
which generates Fortran code calling the HELAS subroutines. At present 
MadGraph is capable of generating the code for the squared matrix elements 
for SM processes with up to seven external particles. It already has been 
used successfully in state of the art calculations like $Z+4$~jets 
production in $p\bar p$ collisions~\cite{BMPS}.

\begin{figure}[htb]
\epsfxsize=5.8in
\epsfysize=0.83in
\begin{center}
\hspace*{0in}
\epsffile{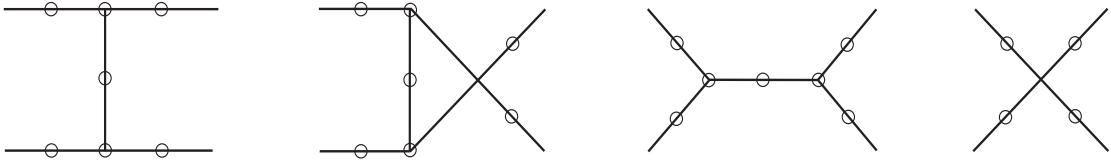}
\caption{
Allowed topologies for Feynman graphs with four external particles. 
The circles indicate the 25 locations where a fifth external particle 
could be attached. 
\label{fig:Feynman-topology}
}
\vspace*{-0.1in}
\end{center}
\end{figure}

Let us consider the various steps which MadGraph takes to generate a program
for the calculation of $\overline{\sum} |{\cal M}|^2$. The first step in 
calculating the scattering amplitude $\cal M$ is to generate 
all possible topologies for tree level Feynman graphs, keeping in mind that 
the SM, as a renormalizable field theory, allows 3 and 4 particle vertices 
only. As an example, the four possible topologies for Feynman graphs with 
four external particles are shown in Fig.~\ref{fig:Feynman-topology}. A fifth
external particle could be attached to each of these graphs in the locations 
indicated by the open circles, leading to 25 distinct tree level topologies 
for Feynman graphs with 5 external particles. For larger numbers of particles 
the number of allowed topologies grows very rapidly, as shown in 
Table~\ref{table:topologies}. This rapid growth leads to increased memory
demands of MadGraph and puts a practical limit of 7 external particles on 
present day workstations. 

\begin{table}[htb]
\caption{Number of distinct topologies\protect\cite{feynarts}
of tree level Feynman graphs 
in a renormalizable field theory as a function of the number 
of external particles.
\label{table:topologies}
}
\vglue0.1in
\tabcolsep=1.47em
\begin{tabular}{l||c|c|c|c|c|c}
\# external particles& 3 & 4 & 5 & 6 & 7 & 8 \\ 
\hline
\# topologies & 1 & 4 & 25 & 220 & 2485 & 34300
\end{tabular}
\end{table}

The tree level amplitude is uniquely fixed once the external particles 
and the interaction lagrangian are specified\footnote{ 
For SM subprocesses involving four or more external quarks both gluon and 
electroweak boson exchange is possible between the quark lines. In these 
cases, MadGraph allows the user to choose the desired order in the 
strong coupling constant $\alpha_s$.}. 
In general only a subset of the allowed topologies will actually
contribute to a given subprocess. Only the third topology in 
Fig.~\ref{fig:Feynman-topology} contributes to $e^+e^- \to \mu^+\mu^-$, 
for example, but with two distinct Feynman graphs, corresponding to photon 
and $Z$ exchange. MadGraph contains a table of all allowed SM vertices. 
For a given topology it first assigns the external particles to the 
external lines in all permutations. Madgraph then works inwards and checks 
at each vertex with one unknown line whether 
there is an allowed assignment of the not yet labeled internal propagator, 
according to the vertex table. Storing the allowed assignments, the process 
is continued until either no allowed particle assignment for the last 
propagator can be found at some vertex or all propagators have been labeled.
In the latter case the complete assignment is stored as a Feynman graph
which needs to be calculated.

The HELAS code generation follows a similar scheme. First the external 
particle wave functions are calculated (i.e. the appropriate calls of 
{\tt IXXXXX, OXXXXX} or {\tt VXXXXX} are printed in the output program). 
The wave functions are, of course, common to all the Feynman graphs. 
For a vertex with one unknown line, one of the function
calls described in the previous subsection then determines the (virtual 
particle) wave function corresponding to this line, be it a current for a 
gauge boson or a 4-spinor for a fermion. This procedure ends at a last vertex
where all wave functions of the incident particles have been calculated.
One final subroutine call then determines the complex value, $A_i$, of the 
amplitude for the given Feynman graph.

The $A_i$ only contain the propagator factors, coupling constants and the 
Lorentz structure of the vertices. For QCD processes the color factors are 
not yet included. For a given Feynman graph the color factor is a product 
of factors from each vertex\footnote{Each 4-gluon vertex leads to
three  distinct Feynman graphs in MadGraph, corresponding to the three 
different color structures in the Feynman rule
\bq
-f^{abe}f^{cde}\left(g_{\mu\rho}g_{\nu\sigma}-g_{\mu\sigma}g_{\nu\rho}\right)
-f^{ace}f^{dbe}\left(g_{\mu\sigma}g_{\nu\rho}-g_{\mu\nu}g_{\sigma\rho}\right)
-f^{ade}f^{bce}\left(g_{\mu\nu}g_{\rho\sigma}-g_{\mu\rho}g_{\nu\sigma}\right)
\nonumber \;. 
\eq
}. These are, for example,
\ba
q_i\;\bar q_j\;g^a\quad {\rm vertex:}&&\qquad 
T^a_{ij}={\lambda^a_{ij}\over 2}\;,  \\
g^a \; g^b \; g^c \quad {\rm vertex:}&&\qquad 
\; i\;f^{abc}\;,   \\
g^a \; g^b \; g^c \; g^d \quad {\rm vertex:}&&\qquad 
-f^{abe}f^{cde}\qquad({\rm 3\; vertices )}\;,  \\
q_i \; \bar q_j\;\gamma\quad {\rm vertex:}&&\qquad 
\delta_{ij}\;.
\ea
In the squared amplitude all colors are summed/averaged over and MadGraph 
uses the familiar commutator and completeness relations
\ba
i\;f^{abc} & = & 2\;{\rm tr} T^a[T^b,T^c]\;,\\
T^a_{ij}T^a_{kl} & = & {1\over 2} \delta_{il}\delta_{kj}-{1\over 2N}
\delta_{ij}\delta_{kl}\;,
\ea
to calculate the color summed/averaged products, $C_{ij}$, of the color 
factors for any pair of Feynman graphs, $i,j$. Thus, the squared amplitude 
takes the form
\bq
\overline{\sum} |{\cal M}|^2= \sum_{i,j=1}^f\; A_i^*\; C_{ij}\; A_j\; .
\eq
A final trick is then used to simplify the color structure\cite{eigencol}:
the hermitian matrix $C_{ij}$ is diagonalized, yielding $n_{max}$ nonzero
eigenvalues $\lambda_n$ 
and orthogonal eigenvectors $z_{nj}$,
\bq
\sum_{j=1}^f C_{ij}z_{nj}^* = \lambda_n \; z_{ni}^*\;.
\eq
This allows to write the squared amplitude in terms of a much smaller
$n_{max}\times f$ dimensional matrix $z$ and a set of $n_{max}$ eigenvalues
$\lambda_n$,
\bq
\overline{\sum} |{\cal M}|^2= \sum_{n=1}^{n_{max}} \lambda_n
|\sum_{i=1}^f\; z_{ni}A_i|^2\; .
\eq

In MadGraph the eigenvalues $\lambda_n$ and the $z_{ni}$ are also used to 
absorb additional factors which appear in the cross section formulas for
specific processes:
\begin{itemize}
\item{}
A relative factor $-1$ between subamplitudes which are obtained by 
interchanging external fermions.
\item{}
A factor $1/N!$ for each set of $N$ identical particles in the final state.
\item{}
An overall factor $(\#\; {\rm initial\; colors})^{-1}$ from taking the 
average over initial state colors.
\end{itemize}
These last two factors must be kept in mind when using the simple crossing 
relations of the HELAS functions to obtain squared amplitudes for crossing
related processes: MadGraph generates the code for calculating the 
polarization and color {\it averaged} amplitude squared for a specific 
input process.

Using MadGraph actually is very simple, one merely needs to type in the 
desired process in a self-explanatory shorthand fashion. It is prudent,
however, to keep the general MadGraph and HELAS properties in mind, so as 
to be able to change the generated programs to ones needs. Desired changes
may involve the addition of new physics contributions to the amplitudes,
a more sophisticated treatment of finite width effects, or a more economic
programming of crossing related processes. One should always keep in mind
that automated systems are merely intended to free us from the labor of
mindless coding. They are tools which make studying the dynamics of 
multi-parton processes more enjoyable and which allow us to concentrate on 
all the physics aspects of signal and background processes.

\section{Manifestations of Color: Rapidity Gaps at the Tevatron
\label{sec:rapgap}}

One important aspect distinguishing signal and background processes is 
the color flow in the contributing Feynman graphs. Many 
signal processes involve electroweak interactions and are thus due to the 
exchange of color singlet quanta while the QCD backgrounds are mediated
by color octet gluon exchange. The question arises whether this different 
color flow has observable consequences which can then be used for background
reduction. There is no general answer to this question. However, there is
an important class of processes, mediated by $t$-channel color singlet
exchange, where the answer is a clear ``yes''. One of the most important 
fields of study at the LHC, weak boson scattering, belongs to this class 
and will be addressed in more detail in the next Section. Let us start 
here with a much simpler example, namely dijet production at the Tevatron, 
which has emerged as an unexpectedly rich QCD laboratory in recent years. 
The discussion will closely follow Ref.~35. 

\begin{figure}[htb]
\epsfxsize=4.5in
\epsfysize=1.5in
\vspace*{-0.1in}
\begin{center}
\hspace*{0in}
\epsffile{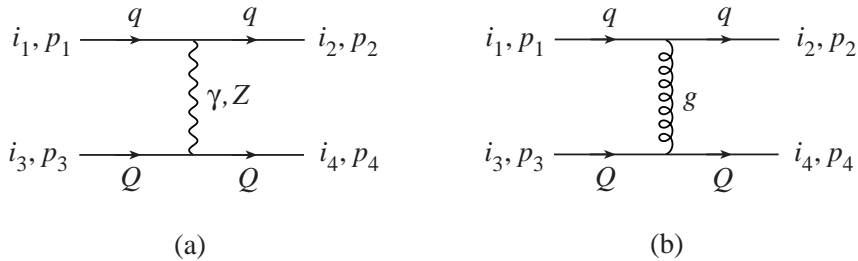}
\vspace*{-0.1in}
\caption{
Feynman graphs contributing to elastic quark-quark scattering via a)
color singlet photon or $Z$ exchange and b) color octet gluon exchange.
\label{fig:qQqQ}
}
\vspace*{-0.1in}
\end{center}
\end{figure}

\subsection{Quark-quark scattering at the tree level}

A number of $2\to 2$ subprocesses contribute to dijet production: gluon-gluon
scattering, quark-gluon scattering, $q\bar q$ annihilation and pair production,
and quark-quark or quark-antiquark elastic scattering. For simplicity let us 
concentrate on the Rutherford scattering process
\bq
q_1(p_1,i_1) Q_3(p_3,i_3) \to q_2(p_2,i_2) Q_4(p_4,i_4)\; ,
\label{eq:qQqQ}
\eq
where the $i_n$ and $p_n$ denote the colors and momenta of the quarks. The two
types of SM contributions, namely color singlet $\gamma/Z$ exchange and color 
octet gluon exchange in the $t$-channel are depicted in Fig.~\ref{fig:qQqQ}. 
Neglecting the $Z$ contribution in the following, the matrix elements for the 
two cases can be written as
\ba
{\cal M}_\gamma & = & \delta_{i_2i_1}\delta_{i_4i_3}\;e_qe_Q\;A =
\delta_{i_2i_1}\delta_{i_4i_3}\;e_qe_Q\; 
{\overline u(p_2)\gamma^\mu u(p_1)\; \overline u(p_4)\gamma_\mu u(p_3)
\over (p_1-p_2)^2 } \; , \\
{\cal M}_g & = & T^a_{i_2i_1}T^a_{i_4i_3}\;g^2\;A = 
\left( {1\over 2} \delta_{i_2i_3}\delta_{i_4i_1}-
{1\over 2N}\delta_{i_2i_1}\delta_{i_4i_3} \right)\; g^2\; A\; .
\label{eq:MgqQqQ}
\ea
The Kronecker deltas in ${\cal M}_\gamma$ describe the fact that the color 
of $q_1$ is directly transferred to $q_2$ while in the QCD case the exchanged 
gluon transmits the color of $q_1$ to $Q_4$ (to leading order in $1/N$). 
Does this difference in color structure lead to observable consequences?

A qualitative understanding of these differences can be obtained by 
considering the acceleration of the color charges in forward scattering, 
at small $|t|=-(p_1-p_2)^2 =s/2\;(1-{\rm cos}\;\theta)<< (p_1+p_3)^2=s$. 
In the case of $t$-channel photon exchange, quark $q$ will be 
deflected by a small angle $\theta$ with respect to the original 
$q_1$ direction and so will the color charge carried by it. The 
acceleration of the color charge by a small angle results in gluon 
bremsstrahlung in the forward direction only and, similarly, the small 
deflection of the color charge carried by $Q$ results in gluon 
radiation in the backward direction, with very little radiation between the 
two scattered quarks. For color octet gluon exchange, on the other hand, the
color charges are exchanged between the two quarks $q$ and $Q$, i.e. they
are deflected by the large angle $\pi -\theta$, and this strong acceleration
of color charges results in strong gluon bremsstrahlung at large angles,
between the directions of the final state quarks.

\begin{figure}[htb]
\epsfxsize=4.5in
\epsfysize=5.5in
\vspace*{-1.5in}
\begin{center}
\hspace*{0.75in}
\epsffile{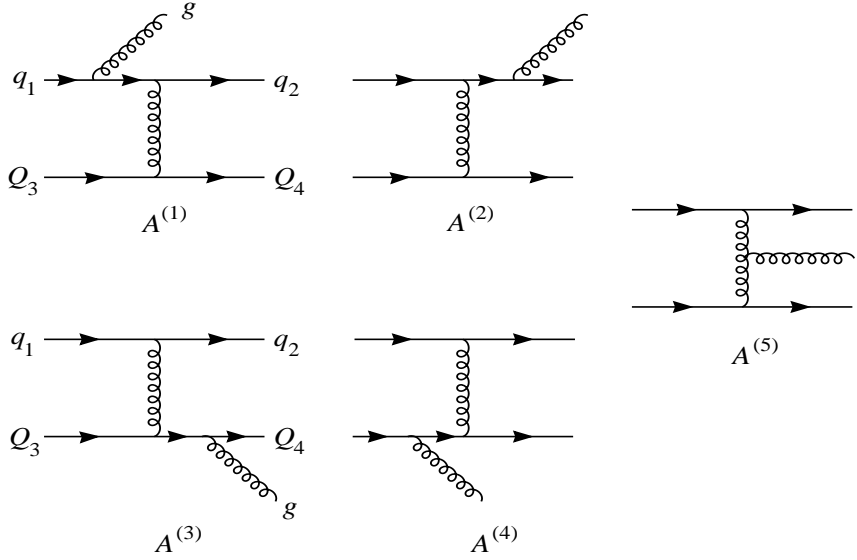}
\vspace*{-1.4in}
\caption{
Feynman graphs for the process $qQ\to qQg$ at tree level.
\label{fig:qQqQg}
}
\vspace*{-0.1in}
\end{center}
\end{figure}

This qualitative picture is confirmed by calculating the gluon emission 
corrections to $qQ$ elastic scattering, i.e. by considering the process 
\begin{equation} \label{process}
q(p_1,i_1)\,Q(p_3,i_3) \longrightarrow q(p_2,i_2)\,Q(p_4,i_4)\,g(k,a)\; ,
\end{equation} 
where $k$ and $a$ denote the gluon momentum and color. At any order of 
perturbation theory, the amplitude for this process can be written in terms 
of four orthogonal color tensors, which we denote by 
$O_{1}$, $O_{2}$, $S_{12}$, and $S_{34}$,
\begin{equation}\label{amp} 
{\cal M} = O_{1}M_{1} + O_{2}M_{2} + S_{12}M_{12} + S_{34}M_{34}\; .
\end{equation} 
In an $SU(N)$ gauge theory they are given explicitly by 
\begin{eqnarray}\label{colortensor}
S_{12} & = & T_{i_4i_3}^{a}\; {\delta_{i_2i_1}} \\
S_{34} & = & T_{i_2i_1}^{a}\; {\delta_{i_4i_3}} \\
O_{1} & = & \frac{-2}{N}\; ( S_{12} + S_{34} ) + 
            T_{i_2i_3}^{a}\; {\delta_{i_4i_1}} + 
            T_{i_4i_1}^{a}\; {\delta_{i_2i_3}} \\  
O_{2} & = & T_{i_4i_1}^{a}\; {\delta_{i_2i_3}} - 
            T_{i_2i_3}^{a}\; {\delta_{i_4i_1}}  \; .
\end{eqnarray} 
$S_{12}$ marks the process where the quark $q$ keeps its color. Similarly, 
$S_{34}$ multiplies the amplitude for $t$-channel color singlet exchange as 
viewed from quark $Q$.  Within QCD, $O_{1}$ and  $O_{2}$ correspond to   
$t$-channel color octet exchange as viewed from either of the two scattering 
quarks. 

Let us first apply this color decomposition to 
the tree level QED and QCD amplitudes. The five Feynman graphs for 
the QCD process are shown in Fig.~\ref{fig:qQqQg}. Lumping the momentum and 
helicity dependence of the individual Feynman diagrams into reduced 
amplitudes $A^{(1)}\cdots A^{(5)}$, one obtains for the QCD amplitudes 
at tree level
\begin{eqnarray}\label{ampQCD}
M^{\rm QCD}_{12} & = &  \frac{-g^3}{2N}\; 
              \left( A^{(1)} + A^{(2)} \right) 
               \equiv  \frac{-g^3}{2N}\; A^{(12)}\\ 
M^{\rm QCD}_{34} & = &  \frac{-g^3}{2N}\; 
              \left( A^{(3)} + A^{(4)}\right)  
               \equiv \frac{-g^3}{2N}\; A^{(34)}\\
M^{\rm QCD}_{1}  & = &  \frac{-g^3}{4}\; 
              \left( A^{(1)} + A^{(2)} + A^{(3)} + A^{(4)}\right) \\ 
M^{\rm QCD}_{2}  & = &  \frac{-g^3}{4}\; 
              \left(A^{(1)} - A^{(2)} + A^{(3)} - A^{(4)} + 2A^{(5)}\right) 
               \equiv \frac{-g^3}{4}\; A^{({\rm na})} \;.
\end{eqnarray} 
The non-abelian three-gluon-vertex only contributes to the color octet 
exchange amplitude $M_2$. Both $M_2$ and $M_1$ vanish for $t$-channel 
photon exchange, while the color singlet exchange amplitudes are given by
\begin{eqnarray}\label{ampQED}
M^{\rm QED}_{12} & = &  -ge_qe_Q\; 
              \left( A^{(3)} + A^{(4)} \right)  
                   = -ge_qe_Q\; A^{(34)}\\ 
M^{\rm QED}_{34} & = &  -ge_qe_Q\; 
              \left( A^{(1)} + A^{(2)}\right)   
                   = -ge_qe_Q\; A^{(12)}\;,
\end{eqnarray} 
where $e_q$ and $e_Q$ denote the electric charges of the two quarks.

\setlength{\unitlength}{0.7mm}
\begin{figure}[htb]
\input rotate
\begin{center}
\vspace*{-0.4in}
\hspace*{0in}
\setbox1\vbox{\epsfysize=6in\epsffile{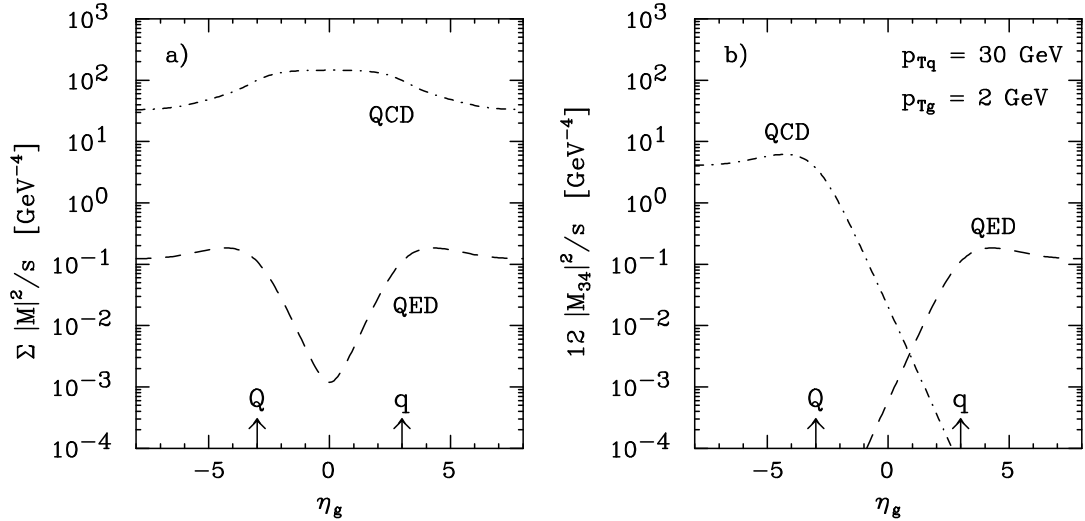}}
\rotl1
\vspace*{-1.6in}
\caption{
Rapidity distribution of emitted gluons in $qQ\to qQg$ scattering for
fixed final state parton transverse momenta of $p_{Tq}=30$~GeV and
$p_{Tg} = 2$~GeV. The quark rapidities are fixed at $\eta_q = \pm 3$
(indicated by the arrows). In part a) results are shown for the sum over all
color structures for single gluon and for $t$-channel photon exchange. 
The $M_{34}$ terms alone, in part b), demonstrate the difference between 
the QED and the QCD color singlet exchange terms. Quark charges are taken as
$e_q=e_Q = +2/3e$.
\label{QEDQCDtree}
}
\vspace*{-0.1in}
\end{center}
\end{figure}

The resulting color and polarization summed squared amplitudes,
\begin{equation}\label{sumM2}
\sum |{\cal M}|^2 = {N^2-1 \over 2}N\, \sum_{\rm polarizations}
\left( |M_{12}|^2 + |M_{34}|^2 
+ 2{N^2-4\over N^2}|M_1|^2 + 2|M_2|^2 \right)\; ,
\end{equation}
are shown in Fig.~\ref{QEDQCDtree}(a) for both the QCD and the QED case. 
Shown is the dependence of $\sum |{\cal M}^2|/s$ on the rapidity of the 
emitted gluon when all other phase space parameters are kept fixed, namely 
the two quarks are held at $p_T=30$~GeV and pseudorapidities $\eta_q=3$ and 
$\eta_Q=-3$ and the gluon transverse momentum is chosen to be 
$p_{Tg}=2$~GeV. The details of this choice are irrelevant: forward 
scattering of the two quarks is a sufficient condition to obtain the 
qualitative radiation pattern of Fig.~\ref{QEDQCDtree}.  
In the QCD case the color octet contributions, via  the non-abelian
amplitude $A^{({\rm na})}$, lead to enhanced gluon emission
in the angular region between the two jets. For $t$-channel photon exchange 
this region is essentially free of gluons due to color coherence 
between initial and final state gluon radiation~\cite{colcoh,fletcher}: 
gluon emission into the central region is exponentially suppressed as the 
rapidity distance from the quarks increases. 

Going beyond the parton level scattering process and considering the actual 
$p\bar p$ scattering event, hadronization of the emitted gluons will result 
in (soft) hadrons which will trace the angular distribution of their parent 
partons. In $t$-channel color singlet exchange, gluon radiation is severely 
suppressed in the angular region between the quarks, and, thus, a similar 
pattern of soft hadrons is expected. This leaves a rapidity gap, i.e. a 
rapidity region between the two quark jets which is essentially void of 
produced hadrons. 

\subsection{Rapidity gaps at the Tevatron\label{sec:rapgapTeV}}

\begin{figure}[htb]
\epsfxsize=3.5in
\epsfysize=3.5in
\vspace*{-0.1in}
\begin{center}
\hspace*{0.0in}
\epsffile{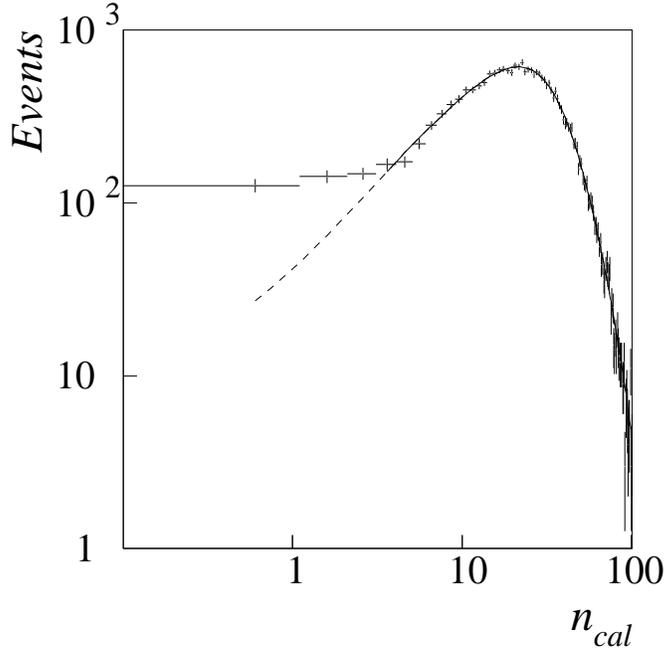}
\vspace*{-0.1in}
\caption{
Electromagnetic tower multiplicity between widely separated jets as observed 
by D0\protect\cite{D0gap2}. Because of the logarithmic $n_{cal}$ scale all 
multiplicities are shifted by +0.5 to the right. The excess of low 
multiplicity (rapidity gap) events above the double negative binomial 
fit (solid and dashed line) is due to color singlet exchange in the 
$t$-channel.
\label{fig:D0gap}
}
\vspace*{-0.1in}
\end{center}
\end{figure}

Rapidity gaps in hard dijet events have indeed been observed at the 
Tevatron\cite{D0gap1,CDFgap,D0gap2}. The D0 Collaboration, for example, has 
studied events with two hard, forward and backward jets with 
\bq
E_{T} > 30\;{\rm GeV}\;,\qquad |\eta_j|>2\;,\qquad \eta_1\cdot \eta_2 < 0\;,
\eq
and has then searched for signs of hadronic activity in the pseudorapidity 
range between the two jets. Inside the region of width $\Delta\eta_c$ between
the tangents to the two jet definition cones (of radius $R=0.7$ in the 
lego-plot) D0 has then studied the hadronic multiplicity in terms of the 
number of active towers in the electromagnetic calorimeter, above a minimum 
transverse energy threshold of 200~MeV. The measured multiplicity 
distribution is shown in Fig.~\ref{fig:D0gap}.
One observes a clear excess of low multiplicity events, above the 
expectation from extrapolating a double negative binomial distribution
which fits very well the high multiplicity region. This excess corresponds to
a fraction 
\bq
f_{gap}= 1.07\pm 0.10{\rm (stat)}^{+0.25}_{-0.13}{\rm (syst)}\;\%
\eq
of rapidity gap events in dijet production. In addition, this fraction is 
essentially independent of $\Delta\eta_c$, for pseudorapidity 
separations larger than 
$\Delta\eta_c\approx 2$ between the jet definition cones\cite{D0gap1}. 
This independence of scattering angle in the parton c.m. frame (for forward 
scattering) implies that the rapidity gap events follow a $1/t^2$ Rutherford
scattering distribution, just like the non-gap dijet events to which they
are normalized and which are dominated by single gluon exchange in the 
$t$-channel.

This angular distribution is exactly as would be expected if the gap events 
were produced by $t$-channel photon exchange. However, the observed rate is 
too large to be consistent with an electroweak origin. The  dijet cross 
section from $t$-channel photon, $Z$ or $W$ exchange is between $10^{-3}$ 
and $10^{-4}$ of the QCD dijet cross section\cite{chehimeetal} in the 
phase space region considered by D0 and, hence, cannot account
for the one percent fraction of dijet events which exhibit a rapidity 
gap. 

Actually, the discrepancy is even larger because a rapidity gap, even if 
present at the level of a single parton collision, may be covered by double 
parton scattering (DPS) or, more generally, the underlying event. The 
hardness of the produced jets translates into an impact parameter between 
the scattering partons which is negligible compared to the transverse size 
of the incident protons. This makes it quite likely that some of the other 
partons in the $p$ and the $\bar p$ will scatter as well, via gluon (color) 
exchange, and the resulting color recombination would lead to the filling 
of the rapidity gap. 
Only the fraction of dijet events without DPS or an underlying event can be 
expected to preserve a rapidity gap. This fraction, the survival probability 
$P_s$ of rapidity gaps, has been estimated by several 
authors\cite{bjgap,fletcher,gotsman} 
for both the Tevatron and supercolliders and it 
is expected to be in the 3--30\% range. For illustration I will use a value 
of $P_s=10$\% in the following. At the LHC $P_s$ can be measured independently,
by comparing the number of electroweak $qq\to qqW$ or $qq\to qqZ$ events 
which exhibit a 
rapidity gap with the SM cross sections for these processes\cite{CZ:Ps}.

\subsection{Two gluon color singlet exchange and rapidity gaps}

\begin{figure}[htb]
\epsfxsize=4.5in
\epsfysize=1.5in
\vspace*{-0.1in}
\begin{center}
\hspace*{0in}
\epsffile{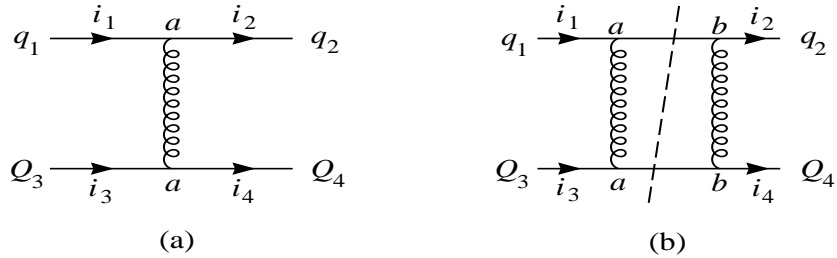}
\vspace*{-0.1in}
\caption{
Feynman graphs for quark scattering via (a) single gluon exchange and 
(b) exchange of two gluons in a color singlet state.
\label{fig:glue-pom}
}
\vspace*{-0.1in}
\end{center}
\end{figure}

Given that electroweak quark scattering cannot account for the observed 
rate of rapidity gap events at the Tevatron, there must be a QCD source of
$t$-channel color singlet exchange. Indeed, the exchange of two gluons in a
color singlet state, as depicted in Fig.~\ref{fig:glue-pom}(b), provides a 
ready source and has been suggested by Low and Nussinov as the basic model 
for the Pomeron\cite{low}.

The rate of rapidity gap events can be understood qualitatively 
in terms of forward scattering by single gluon and two gluon color 
singlet exchange as shown in Fig.~\ref{fig:glue-pom}. The observed rate was, 
in fact, predicted by Bjorken\cite{bjgap}. For forward $qq$ scattering the 
matrix elements for the dominant helicity configurations are given by
\bq
{\cal M}_g \approx 
{8\pi \alpha_s\hat{s}\over Q^2}\;T^a_{i_2i_1}\; T^a_{i_4i_3}
\eq
for single gluon exchange, while color singlet two gluon exchange is 
dominated by the imaginary part of the amplitude in the forward region,
\ba
{\cal M}_{singlet} & \approx &
i\;{8\pi \alpha_s\hat{s}\over Q^2}\;\;{\alpha_s\over 2}{\rm log}(Q^2R^2)\;\;
\left((T^bT^a)_{i_2i_1}\;(T^bT^a)_{i_4i_3}\right)_{singlet}\; \nonumber \\
& = &
i\;{8\pi \alpha_s\hat{s}\over Q^2}\;\;{\alpha_s\over 2}{\rm log}(Q^2R^2)\;\;
{\rm tr}(T^bT^a){\delta_{i_2i_1}\over 3}\;
{\rm tr}(T^bT^a){\delta_{i_4i_3}\over 3}\;.
\ea
Here $R$ is a cutoff parameter, of the order of the proton radius, which is 
needed because the intermediate phase space integral for the Feynman 
graph in Fig.~\ref{fig:glue-pom}(b) is infrared divergent.
The fraction of two gluon color singlet exchange events is then given by
\ba
{\hat\sigma_{singlet}\over\hat\sigma_g} &=& 
{ {4\over 9} |{\rm Im}{\cal M}_{singlet}|^2 \over 2|{\cal M}_g|^2 } 
= {2\over 9}\left| {1\over 2}{\rm log}(Q^2R^2)\;\alpha_s(Q^2)\right|^2\;
\nonumber \\
& \approx & {2\over 9}\left|{1\over 2}{\rm log}(Q^2R^2)\;
{12\pi \over (33-2n_f)\;{\rm log}{Q^2\over \Lambda_{QCD}^2}}\right|^2\;,
\ea
where we have used the LO formula for the running coupling constant at the
scale $Q^2$ which is the physical scale of the process. For sufficiently large
momentum transfer $Q^2=|\hat t|$ the two logs are approximately equal and one 
finds a constant color singlet exchange fraction,
\bq
{\hat\sigma_{singlet}\over\hat\sigma_g}\approx
{1\over 2} \left| {4\pi\over 33-2n_f} \right|^2 \approx 0.15\; ,
\eq
for $n_f=5$.
Thus about 15\% of all quark-quark scattering events are due to color singlet
two gluon exchange and may lead to a rapidity gap, provided that no underlying 
event is present. The probability for the latter is given by the survival
probability $P_s$ and thus the expected fraction of dijet events from 
quark-quark scattering with an 
observable rapidity gap is expected to be\cite{bjgap}
\bq
f_{qq} = {d\sigma_{qq,gap}/dt\over d\sigma_{qq,dijet}/dt}\;
\approx\; 0.15\;P_s\; \approx\; 0.015\;,
\eq
which agrees well with the D0 value of 0.0107. This agreement, however, 
is at least partially by accident. In the forward scattering region, the 
imaginary parts of the color singlet exchange amplitudes in quark-gluon 
and gluon-gluon scattering are the same as the quark-quark scattering 
amplitude, except for a color factor\cite{bjgap},
\bq
{\rm Im}\,{\cal M}_{gg,singlet} ={9\over 4}\,{\rm Im}\,{\cal M}_{qg,singlet}
=\left({9\over 4}\right)^2{\rm Im}\,{\cal M}_{qq,singlet}\;,
\eq
and, at small $t$, the same proportionality is found for the single 
gluon exchange cross sections, 
\bq
{d\sigma_{gg}\over dt}= {9\over 4}\;{d\sigma_{qg}\over dt}
=\left({9\over 4}\right)^2{d\sigma_{qq}\over dt}\;.
\eq
Since the singlet exchange cross section is proportional to 
$({\rm Im}\;{\cal M}_{singlet})^2$, the gap fractions for the three 
channels are related by
\bq
f_{gg} \approx {9\over 4} f_{qg} \approx \left({9\over 4}\right)^2 f_{qq}\; ,
\eq
and, thus, the fraction of color singlet exchange events is substantially 
larger than the 15\% expected for quark-quark scattering alone. 
Using the measured rapidity gap fraction of $\approx 1\%$ this in turn 
indicates a survival probability well below 10\% for the Tevatron.

The rate estimates above assume that two gluon color singlet exchange, i.e. 
the exchange of a Low-Nussinov pomeron, does indeed lead to the same kind 
of gluon radiation pattern as $t$-channel photon exchange and, thus, can 
produce a rapidity gap after hadronization. However, 
the intermediate phase space integral for the Feynman graph of 
Fig.~\ref{fig:glue-pom}(b) is dominated by the phase space region where 
one of the exchanged gluons is very soft. Thus the color of the other, hard
gluon is screened at large distances, i.e. the Low-Nussinov pomeron is
an extended object with colored constituents. As a result, analogy to single 
photon exchange may be misleading: gluon radiation may resolve the internal 
color structure of the pomeron. 
In order to answer this question we need to find out whether 
real gluon emission in Low-Nussinov pomeron exchange follows the pattern
typical for $t$-channel gluon exchange or photon exchange~\cite{CZ}. Only 
in the second case can we expect Low-Nussinov pomeron exchange to lead 
to rapidity gaps.

We already have studied the general color structure of the process
$q_{i_1}Q_{i_3}\to q_{i_2}Q_{i_4}g^a$. The amplitude contains two color octet
pieces $M_1$ and $M_2$ and the color singlet exchange amplitudes $M_{12}$ and 
$M_{34}$. Let us concentrate on the last one, which describes color singlet 
exchange as seen by quark $Q$.
Even for $t$-channel gluon exchange does this color singlet amplitude exist. 
However, the rapidity distribution of the emitted gluon is markedly different 
from photon exchange. 

In the QED case the $M_{34}$ amplitude corresponds to emission of the 
final state gluon off the quark $q$ ($A^{(1)}$ and $A^{(2)}$ in 
Fig.~\ref{fig:qQqQg}). In forward scattering ($\eta_q=+3$ in 
Fig.~\ref{QEDQCDtree}(b)) the gluon is radiated between the initial and 
final state $q$ directions, i.e. at $\eta_g \gsim \eta_q$. 
The color $i_1$ of the initial quark $q$ is 
thus transferred to a low mass color triplet object which emerges close to 
the beam direction. At lowest order this is the final state $q$, at 
${\cal O}(\alpha_s)$ it is the $qg$ system. The situation is thus stable 
against gluon emission at even higher order for the QED case and gluon 
radiation is suppressed in the rapidity range between the two final state 
quarks. 

\begin{figure}[htb]
\epsfxsize=4.3in
\epsfysize=1.4in
\vspace*{-0.2in}
\begin{center}
\hspace*{0.0in}
\epsffile{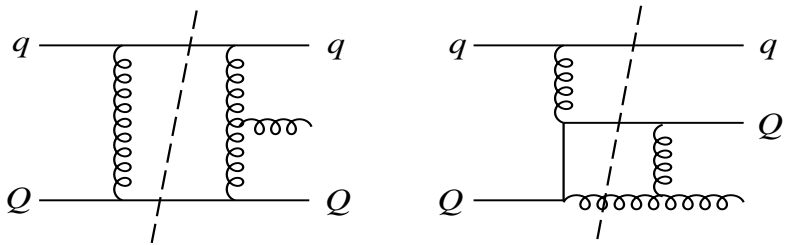}
\vspace*{-0.1in}
\caption{
Two of the 31 Feynman graphs contributing to the imaginary part of 
the color singlet exchange amplitude $M_{34}$.    
\label{fig:pomrad2}
}
\vspace*{-0.1in}
\end{center}
\end{figure}

In the QCD case $M_{34}$ corresponds to emission of the gluon from the quark 
$Q$. The gluon is preferentially emitted between the initial and the final 
$Q$ directions, at $\eta_g\lsim\eta_Q$ (dash-dotted line in 
Fig.~\ref{QEDQCDtree}(b)). Thus the color triplet $qg$ system, into which the 
initial quark $q$ evolves, consists of a widely separated quark and gluon. 
Higher order corrections will lead to strong gluon radiation into the angular 
region between the two and thus also into the rapidity range between the 
two final state quarks. 

These typical patterns found for $t$-channel color singlet and color octet 
exchange can now be used as a gauge for the radiation pattern produced in
$qQ\to qQg$ scattering via the exchange of two gluons in a color singlet state.
In the lowest order process, $qQ\to qQ$, the color singlet exchange amplitude 
is dominated by its imaginary part~\cite{cudell}. Hence, we may estimate the 
radiation pattern 
by calculating the imaginary part of the gluon emission amplitude $M_{34}$ 
only. Typical Feynman graphs are shown in Fig.~\ref{fig:pomrad2}. 
Details of the calculation are given in Ref.~35. 

For massless internal gluon propagators the phase space integrals over the 
$qQ$, $qg$, and $gQ$ intermediate states are divergent. They can be 
regularized by replacing the massless gluon propagator by a version which 
avoids unphysical gluon propagation over long distances~\cite{lan}. QCD 
Pomeron models of this kind have been found to give a good description of 
available data~\cite{natale}. These refinements are approximated by using 
an effective gluon mass of $m_r = 300$~MeV in the calculation.

\begin{figure}[t]
\begin{center}
\input rotate
\vspace*{-0.5in}
\hspace*{0in}
\setbox1\vbox{\epsfysize=6in\epsffile{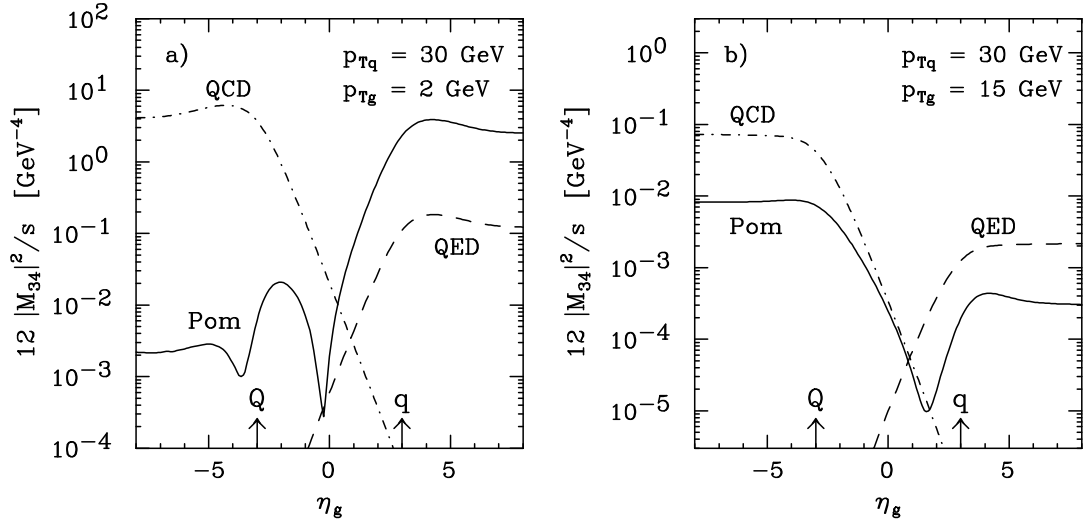}}
\rotl1
\vspace*{-1.6in}
\caption{
Rapidity distribution of emitted gluons in $uc\to ucg$ scattering
via two gluon color singlet exchange as seen by the charm quark.
The phase space parameters for the quarks are the same as in 
Fig.~\protect\ref{QEDQCDtree} and results are shown 
for a) the case of a soft gluon ($p_{Tg}=2$~GeV) and b) a hard gluon 
($p_{Tg}=15$~GeV). For comparison tree level results are shown for gluon 
(dash-dotted lines) and photon exchange (dashed lines).  
 \label{figthree}
}
\vspace*{-0.2in}
\end{center}
\end{figure}

A second problem arises because some of the contributions to ${\rm Im}M_{34}$ 
correspond to $q\to g$ splitting and subsequent $gQ\to gQ$ scattering via 
pomeron exchange. These contributions cannot be expected to be suppressed 
when the gluon is emitted between the $q$ and the $Q$ directions and thus 
would mask the radiation off pomeron exchange in $qQ$ scattering. These 
splitting contributions have been subtracted in Ref.~35 
to yield the square of the pomeron exchange radiation pattern, 
$|{\rm Im}M_{34}^{\rm pom}|^2$, which is shown in Fig.~\ref{figthree}.

\setlength{\unitlength}{0.7mm}

For high transverse momentum of the emitted gluon (of order of the quark 
momenta, see Fig.~\ref{figthree}(b)) the radiation pattern is quite similar 
to the one obtained for single gluon exchange.  Hard 
emitted gluons have too short a wavelength to see the screening of the 
color charge of the harder exchanged gluon by the second, typically very 
soft, exchanged gluon. The Low-Nussinov pomeron thus reveals itself as an 
extended object. 
Hard gluon emission is able to resolve the internal color structure of the 
pomeron.
As the transverse momentum of the emitted gluon is decreased, a qualitative 
transition occurs, as is apparent by comparing the $p_{Tg}=2$~GeV and 15~GeV 
cases in Fig.~\ref{figthree}. The gluon radiation has too long a wavelength to
resolve the internal color structure and hence the pomeron appears as a color 
singlet object. 

In dijet production at the Tevatron, in the kinematic range studied by D0,
gluon radiation will be dominated by the soft region, with transverse gluon 
momenta in the few GeV range or below, as shown in Fig.~\ref{figthree}(a). 
The emission of these soft gluons follows a pattern very similar to the one 
observed for $t$-channel photon exchange, which leads to the formation of 
rapidity gaps. One concludes that two gluon 
color singlet exchange in dijet events does indeed lead to the formation of 
rapidity gap events as observed at the Tevatron.

\section{Weak Boson Scattering \label{sec:qqtoqqVV}}

One of the major reasons for building a hadron supercollider is 
the study of weak boson interactions in the TeV range. Within the forseeable 
future this task will have to be performed in $pp$ collisions at 14~TeV at 
the Large Hadron Collider (LHC) at CERN. Because of its limited energy 
the study of weak boson scattering will be a very demanding task at the 
LHC, requiring a full arsenal of tools to isolate $qq\to qqVV$ weak 
boson scattering events from the backgrounds. For example, jets will 
have to be used as a tagging device (as in the top quark search at 
the Tevatron) and rapidity gap techniques, i.e. the exploitation of 
the $t$-channel color singlet exchange in the signal, should prove 
useful as well, thus combining the ideas discussed in the previous Sections. 
Before considering the benefits of these techniques in weak boson 
scattering and the search for a heavy Higgs at the LHC, let us start with 
a brief note on the physics of a strongly interacting Higgs sector.

\subsection{Higgs production and longitudinal weak boson 
scattering\protect\cite{GHKD}}

In a more general context, the search for the Higgs boson is part of
the quest for the basic interactions which are responsible for the 
spontaneous breaking of the $SU(2)\times U(1)$ gauge symmetry. The existence 
of this local symmetry is evidenced by the observation of $W$ and $Z$ bosons
and the fact that their experimentally determined couplings to quarks and 
leptons agree with the gauge theory predictions. The appearance of gauge 
boson masses then requires the gauge symmetry to be broken spontaneously, 
with some order parameter $\Phi$ acquiring a vacuum expectation value (v.e.v.). 
The fact that the $W$ to $Z$ mass ratio leads to a $\rho$-parameter very 
close to unity finally tells us that this order parameter must
transform essentially as an $SU(2)$-doublet. 

Beyond these basic facts the precise nature of the order parameter and of the 
interactions which drive its v.e.v. must be determined experimentally. Within
the SM the order parameter is simply the scalar Higgs doublet field,
\begin{eqnarray}
\Phi = \left(
\begin{array}{c} \frac{1}{\sqrt{2}}\left(v+H+i\chi^3\right)\\
 i\chi^-
 \end{array} \right), \label{phi}
\end{eqnarray}
where $v$ is the Higgs v.e.v., $\chi^\pm$ and $\chi^0$ are the Goldstone 
bosons which are closely related to the longitudinal degrees of freedom of 
the $W^\pm$ and the $Z$, and $H$ is the field describing the physical Higgs 
boson which we would like to detect experimentally. 

\begin{figure}[t]
\epsfxsize=4.3in
\epsfysize=1.4in
\vspace*{-0.2in}
\begin{center}
\hspace*{0.0in}
\epsffile{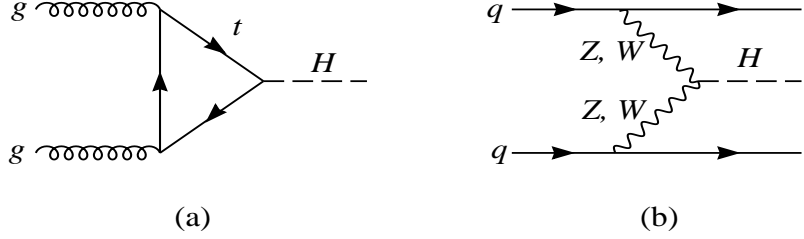}
\vspace*{-0.1in}
\caption{
Feynman graphs for the two dominant Higgs production processes at the LHC,
(a) gluon-gluon fusion via a top-quark loop and (b) weak boson fusion.
\label{fig:Hprodfeyn}
}
\vspace*{-0.1in}
\end{center}
\end{figure}

Without spontaneous breakdown of the symmetry, $SU(2)\times U(1)$ gauge 
invariance forbids any mass terms for the known quarks and leptons as well 
as for the gauge bosons. Thus these masses must be proportional to the v.e.v. 
Since $v$ appears only in the combination $v+H$ in the SM 
(see Eq.~(\ref{phi})), this implies couplings of the fermions and gauge 
bosons to the Higgs which are proportional to the masses of the former. Let 
us consider the most important of these, the couplings of the Higgs to the 
gauge bosons and to the top quark. They are derived from the kinetic 
energy term for the Higgs doublet field and from the Yukawa coupling of 
the Higgs field to the top-bottom doublet,
$T=(t_L,b_L)^T$, and the right handed top quark field, $t_R$,
\ba
{\cal L} & = & (D_\mu\Phi)^\dagger (D^\mu\Phi)\;  - \;
(\lambda_t\; \overline{T} \Phi\; t_R + {\rm h.c.})  \nonumber \\
& = & \left(1+{H\over v}\right)^2\;\left|{i\over 2}\left(
\begin{array}{cc}
g\;W_\mu^3-g'\;B_\mu & \dots \\ 
g\sqrt{2}W_\mu^- & \dots 
\end{array} \right) \left(
\begin{array}{c}
{v\over\sqrt{2}} \\ 0 
\end{array} \right) \right|^2  -
\left(1+{H\over v}\right)\lambda_t\; \overline{t}_L{v\over\sqrt{2}}\;t_R\;
+ \dots\nonumber \\
& = &  \left(1+{H\over v}\right)^2\; \left(m_W^2\;W_\mu^\dagger W^\mu
+ {m_Z^2\over 2}Z_\mu Z^\mu\right) - \;
m_t\; \overline{t}_Lt_R \left(1+{H\over v}\right)\;+\;\dots\;,
\label{eq:HVV}
\ea
which gives relations between the masses and coupling constants,
\bq
m_W = {gv\over 2}\;,\qquad 
m_Z = {gv\over 2\;{\rm cos}\;\theta_W}\;,\qquad 
m_t = {\lambda_t v\over \sqrt{2}}\; .
\label{eq:massrel}
\eq

The interaction terms of Eq.~(\ref{eq:HVV}) imply that the Higgs boson decays
predominantly into the heaviest particles available. For the heavy Higgs 
scenario, which is considered in the following, these are the decay modes
$H\to W^+W^-$, $H\to ZZ$, and $H\to \overline{t}t$. For a Higgs mass of 
800~GeV, for example, the corresponding branching ratios are
\bq
B(H\to W^+W^-) \approx 0.59\;, \qquad
B(H\to ZZ) \approx 0.29\;, \qquad
B(H\to \overline{t}t) \approx 0.12\;,
\eq
and the total Higgs decay width is expected to be about 290~GeV. In order
to exploit the dominant decay modes, one needs to search for the 
Higgs signal in $W^+W^-$ and $ZZ$ production events at the LHC,
with subsequent leptonic decays of at least one of the weak bosons.

\begin{figure}[t]
\epsfxsize=4.0in
\epsfysize=4.0in
\vspace*{0.2in}
\begin{center}
\hspace*{-0.3in}
\epsffile{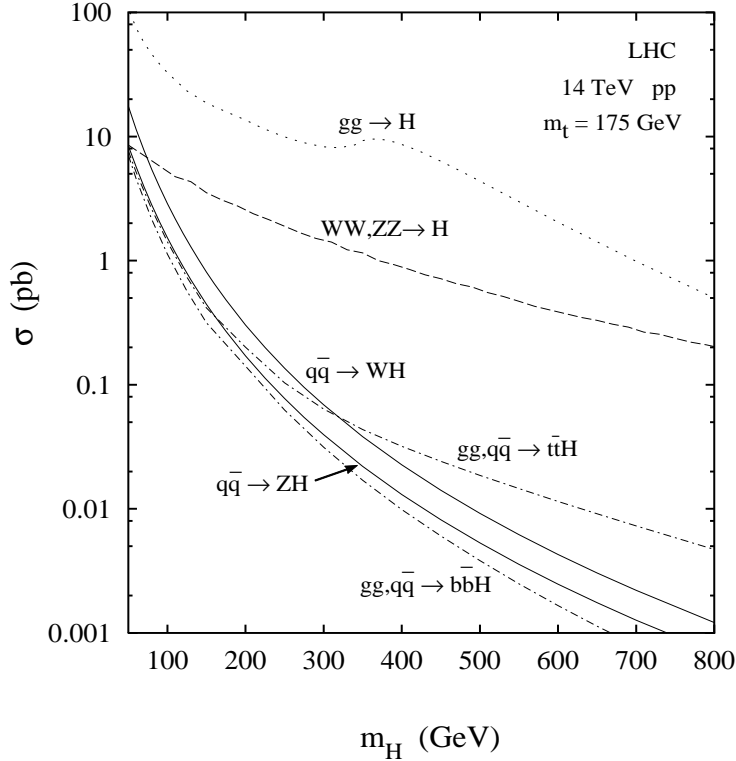}
\vspace*{-0.25in}
\caption{
Production cross sections for the SM Higgs boson at the LHC. 
{}From Ref.~46. 
\label{fig:Hprodcross}
}
\vspace*{-0.1in}
\end{center}
\end{figure}

Similar to the Higgs decay modes, the dominant production processes involve 
heavy particles. At LHC energy, the Higgs boson is mainly produced via gluon 
fusion, $gg\to H$, which proceeds via a top quark loop and is depicted in 
Fig.~\ref{fig:Hprodfeyn}(a), and via weak boson fusion, $WW,\;ZZ\to H$, 
which, more precisely, is the electroweak process $qq\to qqH$ (see 
Fig.~\ref{fig:Hprodfeyn}(b)).
The corresponding production cross sections are shown in 
Fig.~\ref{fig:Hprodcross}. 

While the gluon fusion cross section ranges from 30 to 0.5~pb for 
100~GeV~$<m_H<800$~GeV, weak boson fusion rates are typically smaller by a 
factor 2.5 to 10, the difference becoming smallest at large Higgs boson 
masses. This somewhat smaller cross section, however, is compensated by a 
number of features which are special to the weak boson scattering process.
First of all the additional two final state quarks will frequently manifest
themselves as hadronic jets which allows to suppress backgrounds by jet 
tagging. Second the two production processes probe different interactions
of the Higgs boson. The gluon fusion cross section is proportional to
$\lambda_t^2$, i.e. it measures the Yukawa coupling of the produced scalar to 
the top quark. This Yukawa coupling, however, does not identify the produced 
particle as being connected to the symmetry breaking mechanism, it is 
possible for a generic scalar particle. The $HVV$ coupling, on the other 
hand, at tree level, is only possible 
if $H$ is indeed the Higgs boson. The weak boson fusion rate directly
measures the $HVV$ coupling (as does the $H\to VV$ decay width) and thus
one would like to measure both gluon and weak boson fusion cross sections 
in order to disentangle the various couplings of $H$ in Eq.~(\ref{eq:HVV}).

\begin{figure}[t]
\epsfxsize=4.8in
\epsfysize=1.4in
\vspace*{-0.2in}
\begin{center}
\hspace*{0.0in}
\epsffile{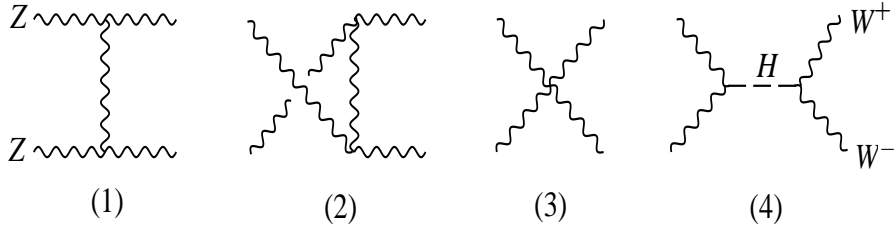}
\vspace*{-0.1in}
\caption{
Feynman graphs for the weak boson scattering process $ZZ\to W^+W^-$. 
\label{fig:VVVV}
}
\vspace*{-0.1in}
\end{center}
\end{figure}

When considering the full processes $qq\to qqWW$ or $qq\to qqZZ$, the 
resonance approximation, i.e. considering graph (b) in Fig.~\ref{fig:Hprodfeyn}
only, is only valid for a Higgs width $\Gamma_H<<m_H$. For a heavy (and wide)
Higgs resonance the $VV\to H(\to VV)$ subprocess in Fig.~\ref{fig:Hprodfeyn}(b)
must be replaced by the full weak boson scattering amplitude, as depicted in 
Fig.~\ref{fig:VVVV} for the process $ZZ\to WW$. In addition, weak boson 
bremsstrahlung off the quark lines must be considered\cite{qqWW,BCHZ}. 

Very strong cancellations between the various Feynman graphs of 
Fig.~\ref{fig:VVVV} appear when all the external vector bosons are 
longitudinally polarized\cite{unitarityH}. The isospin 
and angular momentum zero contributions from the first two graphs and from the 
third graph individually grow as ${\hat s}^2/m_W^4$. However, the leading 
terms cancel when the sum of the three graphs is considered. Even the 
${\hat s}/m_W^2$ growth of their sum, however, would lead to a violation 
of partial wave unitarity at a $\sqrt{\hat s}$ of about 1~TeV. At high 
energies, 
above ${\hat s}=m_H^2$, it is the destructive interference with the fourth 
graph, $s$-channel Higgs exchange, which leads to an acceptable $J=I=0$
partial wave amplitude. Thus the existence of the Higgs boson, or some other,
additional contribution to the weak boson scattering amplitude, beyond the
vector boson self-coupling graphs (1)--(3) in Fig.~\ref{fig:VVVV}, is required
by unitarity of the $S$-matrix\cite{unitarityH}. 
This additional contribution does not have to 
be the $s$-channel exchange of a scalar object. It might be the $t$- or 
$u$-channel exchange of a vector particle which resides in a weak isospin 
triplet, like in technicolor models. Or nature may have realized the 
unitarization of the weak boson scattering amplitudes in yet another 
manner\cite{bagger}.

Depending on which path is realized in nature, the various weak boson 
scattering processes like $W^+W^- \to W^+W^-$, $WZ\to WZ$, $W^+W^+\to W^+W^+$,
etc. will exhibit markedly different magnitude and shape (energy dependence) 
of the cross sections. If a technirho existed, the $WZ$ channel, for example,
would exhibit a resonance at the technirho mass, while this channel has a 
small cross section if a $J=I=0$ scalar is responsible for the unitarization 
of the $S$-matrix (as in the SM). Thus the study of all weak boson scattering
channels is important in order to fully probe the symmetry breaking sector.
In the following I will use the search for a SM heavy Higgs to discuss 
techniques of background suppression. One should keep in mind, however, that
these methods are more general and can be applied to study any of the weak
boson scattering processes.

\subsection{Forward jet-tagging}

A characteristic feature of weak boson scattering events are the two 
accompanying quarks (or antiquarks) from which the ``incoming'' $W$s or 
$Z$s have been radiated (see Fig.~\ref{fig:Hprodfeyn}(b)). In general these 
scattered quarks will give rise to hadronic jets. By tagging them, i.e. by 
requiring that they are observed in the detector, one hopes to 
obtain a powerful background rejection tool\cite{Cahn,Froid}. Whether such an 
approach can be successful depends on the properties of the tagging jets:
their typical transverse momenta, their energies, and their angular 
distributions.

Similar to the emission of virtual photons from a high energy electron 
beam, the incoming weak bosons tend to carry a small fraction of 
the incoming parton energy\cite{equivW}. At the same time the incoming 
weak bosons must carry substantial energy, of order $m_H/2=m_{VV}/2$, in 
order to produce a weak boson pair of large invariant mass. Thus the final 
state quarks in $qq\to qqVV$ events will carry very high energies, of order 
1~TeV or even higher. This is to be contrasted with their transverse momenta, 
which are of order $p_T\approx m_W$. This low scale arises because the weak 
boson propagators in Fig.~\ref{fig:Hprodfeyn}(b) introduce a factor
\bq
D_V(q^2) = {-1\over q^2 - m_V^2} \approx {1\over p_T^2+m_V^2}
\eq
into the production amplitudes and
suppress the $qq\to qqH$ cross section for quark transverse momenta above
$m_V$. The modest transverse momentum and high energy of the scattered quark 
corresponds to a small scattering angle, typically in the $1.5<\eta<4.5$ 
pseudorapidity region. 

\setlength{\unitlength}{0.7mm}
\begin{figure}[htb]
\begin{center}
\input rotate
\vspace*{-0.8in}
\hspace*{0.0in}
\setbox1\vbox{\epsfysize=6.5in\epsffile{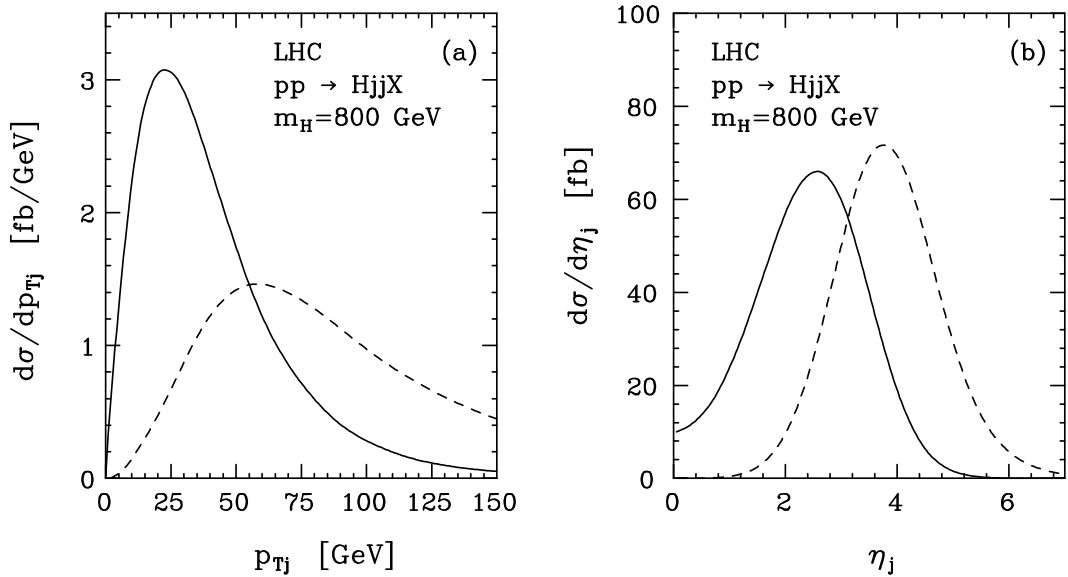}}
\rotl1
\vspace*{-1.2in}
\caption{
Transverse momentum and pseudorapidity distributions of the two (anti)quark 
jets in $qq\to qqH$ events at LHC energy. Shown are (a) $d\sigma/dp_{Tj}$ for
the highest (dashed curve) and lowest $p_T$ jet (solid curve) and (b) 
$d\sigma/d|\eta_j|$ for the most forward (dashed curve) and the most central
jet (solid curve).
\label{fig:ptjH}
}
\vspace*{-0.1in}
\end{center}
\end{figure}

These general arguments are confirmed by Fig.~\ref{fig:ptjH}, where the 
transverse momentum and pseudorapidity distributions of the two potential 
tagging jets are shown for the production of a $m_H=800$~GeV Higgs boson at 
the LHC. One finds that one of the two quark jets has substantially lower 
median $p_T$ ($\approx 30$~GeV) than the other ($\approx 80$~GeV).
As a result, double
jet-tagging, i.e. the requirement that both scattered quarks are visible as 
jets, proves quite costly\cite{Froid,BCHP,BCHZ} unless jets with 
transverse momenta around 30~GeV can be identified in the forward 
region. A different approach is single forward jet-tagging which relies only 
on the higher $p_T$ tagging-jet and thus proves more effective for higher 
transverse momentum thresholds\cite{BCHP,BCHZ,BCHSZ,DGOV}. 

This transverse momentum threshold needs to be set such that the probability
for seeing a fake tagging-jet becomes small. At full LHC luminosity of
${\cal L}=10^{34}{\rm cm}^{-2}{\rm sec}^{-1}$ one of the most important 
sources of such background jets are additional $pp\to jjX$ events in the 
same bunch crossing as the $pp\to VVX$ signal event. With a bunch crossing 
happening every 25~nsec, a process with a cross section of 
$\sigma_{eff} = [{\cal L}\; 25{\rm nsec}]^{-1} = 4$~mb will, on average, happen
during each bunch crossing. Using Poisson statistics, the probability to get 
a random jet of transverse momentum $p_{Tj}>p_{T,\rm cut}$ in any event 
previously selected is then given by
\bq
P_j(p_{T,\rm cut}) = 1-{\rm exp}\left(-{\sigma_{jj}(p_{Tj}>p_{T,\rm cut}) 
\over \sigma_{eff}}\right) \approx {\sigma_{jj}(p_{Tj}>p_{T,\rm cut}) 
\over \sigma_{eff}}\; .
\eq
With single-jet cross sections of about 0.5--1~mb above $p_{Tj}=20$~GeV in the 
relevant rapidity range, overlapping events in a single bunch crossing 
may produce fake tagging-jets above 20~GeV $p_T$ with about 15--25\% 
probability. This estimate agrees with the results of a more 
detailed analysis of overlapping events at the LHC~\cite{ciapetta}.	

In the following let us sidestep this question of how low a transverse momentum
threshold can be achieved in forward jet-tagging. By restricting ourselves to 
single forward jet-tagging the jet $p_T$ threshold can be set high enough
to avoid the fake jet problem from overlapping events. As an example we will
consider the search for $H\to W^+W^-\to \ell^+\nu\ell^-\bar\nu$ decays as
discussed in Ref.~18. 
This particular decay mode suffers from severe
backgrounds due to $t\bar t \to W^+W^-b\bar b$ decays as well as 
$q\bar q\to W^+W^-$ events (referred to as $WW$ QCD background in the 
following) and thus is well suited to assess the power of jet-tagging and 
minijet veto techniques. 

Since we are interested in the decay of a very heavy Higgs boson, the two 
charged $W$ decay leptons will emerge with high transverse momentum, in the 
central region of the detector, and they will be well isolated from additional
jets. Thus we require the presence of two charged leptons ($\ell=e,\mu$) 
with
\begin{equation}\label{cut1a}
p_{T\ell}  >  50\; {\rm GeV}\;, \qquad |\eta_\ell|  <  2\; , \qquad
R_{\ell j} = \sqrt{(\eta_\ell-\eta_j)^2 + (\phi_\ell-\phi_j)^2}  >  0.7\;.
\end{equation}
Here $p_{T\ell}$ denotes the lepton transverse momentum and $\eta_\ell$
is its pseudorapidity. The $R_{\ell j}>0.7$ separation cut forbids a parton 
(jet) of $p_T>20$~GeV in a cone of radius 0.7 around the lepton direction. 

The lepton $p_T$ cut in Eq.~(\ref{cut1a}) is not in itself sufficient to focus 
on the production of two $W$'s of large transverse momenta and large
$W$-pair invariant mass. A variable 
which helps to substantially suppress $W$ bremsstrahlung backgrounds is
$\Delta p_{T\ell\ell}$, the difference of the charged lepton transverse 
momentum vectors~\cite{DGOV}. We thus require
\begin{equation}\label{cut1b}
\Delta p_{T\ell\ell} = | {\bf p}_{T\ell_1}-{\bf p}_{T\ell_2}|>300\, 
{\rm GeV}\;, \qquad m_{\ell\ell}> 200\, GeV\;.
\end{equation}
The additional cut on the dilepton invariant mass removes possible 
backgrounds from $Z$ leptonic decays. It is largely superceded by the 
the $\Delta p_{T\ell\ell}$ cut, however. 

Cross sections for events satisfying the lepton acceptance criteria of 
Eqs.~(\ref{cut1a},\ref{cut1b}) are listed in the first column of 
Table~\ref{tab:Hcross} for the case of a $m_H=800$~GeV Higgs boson and the 
$q\bar q \to W^+W^-$ and $t\bar t$ production backgrounds. In the calculation 
of the $qq\to qqWW$ signal all $W$ bremsstrahlung graphs have been 
included\cite{qqWW,BCHZ} since the narrow Higgs width approximation is
not appropriate any more for the large masses considered here. These additional
graphs introduce an ``electroweak background'' which is the same as in the 
case of a light Higgs boson. Since a light Higgs cannot decay via $H\to WW$, 
we can use the SM $qq\to qqWW$ cross section for, say, $m_H=100$~GeV as an 
estimate of this electroweak background. The heavy Higgs signal is then 
defined as $B\sigma_{\rm SIG} =B\sigma(m_H) - B\sigma(m_H=100\; {\rm GeV})$. 

\begin{table}
\caption{Signal and background cross sections $B\sigma$ (in fb) for the
Higgs search in $qq\to qqWW,\;WW\to \ell\nu\ell\nu$ events. Results are shown
for increasingly stringent cuts. From Ref.~18. 
\label{tab:Hcross} }
\vspace*{0.1in}
\begin{tabular}{lcccc}
& lepton cuts only& + tagging jet&
{\def\arraystretch{.66}
\begin{tabular}[t]{c}
+ lepton-\\
tagging jet\\
separation
\end{tabular}} 
& {\def\arraystretch{.66}
\begin{tabular}[t]{c}
+ minijet veto\\
($p_{T,\rm veto}=$\\
20~GeV)
\end{tabular}}\\
& [Eq.~(\ref{cut1a})--(\ref{cut1b})]& [Eq.~(\ref{cut2})]& [Eq.~(\ref{cut3})]&
[Eq.~(\ref{cutveto})]\\
\hline
$WW(jj)$& 27.4& 1.73& 0.57& 0.13\\
$t\bar t(jj)$& 640& 57& 25& 0.47\\
$m_H=100$ GeV& 1.18& 0.56& 0.29& 0.18\\
$m_H=800$ GeV& 3.4& 1.79& 1.31& 0.97\\[.2in]
\underline{signal}:&&&&\\
$m_H=600$ GeV&&&&0.78\\
$m_H=800$ GeV& 2.2& 1.23& 1.02& 0.79\\
$m_H=1$ TeV&&&& 0.62
\end{tabular}
\end{table}

Without any jet-tagging the backgrounds are overwhelming: the top quark 
background alone exceeds the signal by a factor of 300. How much does this 
situation improve by forward jet-tagging? A large fraction of the signal 
(55\%) is retained by requiring the existence of a very energetic, forward 
jet of moderate $p_T$, in the phase space region 
\begin{equation}\label{cut2}
E_j^{\rm tag} > 500\, {\rm GeV}\;, 
\qquad 1.5 <|\eta_j^{\rm tag}| < 4.5\;,
\qquad p_{Tj}^{\rm tag} > 50\, {\rm GeV} \;.
\end{equation}
In addition the tagging-jet should  be well separated from the $W$ decay 
leptons,
\begin{equation}\label{cut3}
{\rm min}\; |\eta_j^{\rm tag}-\eta_\ell | > 1.7\; ,
\end{equation}

Since we are now requiring additional jet activity, the previous background
calculations are not sufficient any more. Rather ${\cal O}(\alpha_s)$ real 
emission corrections must be considered, i.e. $t\bar t j$ production and
processes like $q\bar q\to W^+W^-g$ and $qg\to W^+W^-q$ need to be 
considered at this level.  
The signal and background cross sections after the cuts of Eqs.~(\ref{cut2}) 
and (\ref{cut3}) are listed in the second and third columns of 
Table~\ref{tab:Hcross}, respectively.  Single forward jet-tagging has 
reduced the backgrounds by a factor 25--50, while keeping 45\% of the signal.
This reduction factor is almost sufficient for the $WW$ QCD background, but
an additional large suppression factor is needed for the top quark decays.
Fortunately, the $b$-quarks in $t\to bW$ decay frequently give rise to 
additional jets and a veto on them\cite{BCHP}, above $p_{Tb}=25$~GeV, will 
substantially reduce this background\cite{BCHZ}. One needs to be careful,
however, since at such low transverse momenta the production of 
minijets via the emission of additional gluons cannot be neglected at the LHC.

\subsection{Rapidity gaps and minijet veto\label{sec:minijveto}}

The most rigorous veto on $b$-quarks from top decays would be a rapidity 
gap requirement\cite{troyan,bjgap} as discussed for dijet events at the 
Tevatron in Section~\ref{sec:rapgapTeV}. A rapidity gap trigger, i.e. the
selection of events without any hadronic activity in some rapidity range 
around the two charged leptons which arise from the two $W$s, would make 
maximal use of the different color structure of signal and background 
processes. In a weak boson scattering event no color is exchanged between 
the initial state quarks. Color coherence between initial and final state
gluon bremsstrahlung then leads to a suppression of hadron production in the 
central region, between the two scattered quarks~\cite{troyan,bjgap,Kane}. 
The $t\bar t$ production and $WW$ QCD backgrounds, on the other hand, involve 
color exchange between the incident partons and, as a result, gluon radiation 
into the central region dominates. This gluon radiation then results in 
substantial hadronic activity in the vicinity of the $W$ decay leptons.

While a rapidity gap trigger should suppress the backgrounds well below the 
signal level, it is not practical at the LHC, due to the small signal rate
and small survival probability of the signal. In order for a rapidity gap 
to be visible no additional scattering process may occur in the detector
at the same time, neither in the same $pp$ collision which produces the 
Higgs boson, nor by other proton pairs which collide in the same bunch 
crossing. The first condition is parametrized by the survival probability,
$P_s$, which was discussed in Section~\ref{sec:rapgapTeV}, and which we expect
to be below 10\% at the LHC\cite{bjgap,fletcher,gotsman}. The second condition
implies that one must run the LHC well below design luminosity, at an average
of one $pp$ scattering event per bunch crossing, or about 
${\cal L}=5\cdot 10^{32}{\rm cm}^{-2}{\rm sec}^{-1}$. Even at this low 
luminosity only $P_1 = e^{-1}=37\%$ of all bunch crossings have exactly one 
$pp$ scattering event, leading to an effective yearly integrated luminosity
of about 2~fb$^{-1}$. Given the signal cross section of 1.02~fb
in the third column of Table~\ref{tab:Hcross} and an (optimistic) survival 
probability of $P_s=10\%$, one would register only one Higgs signal event
every five years! Note that the cuts imposed on the leptons and tagging-jet
are not responsible for this negative result: 22\% of all signal events are 
accepted by them\cite{bpz}. It is the combination of small survival probability
and additional effective luminosity reduction 
by a factor 50 which kills the signal. 
Given the small weak boson scattering cross section, any rapidity 
gap strategy at the LHC must raise this effective survival probability from
1/500 to something of order unity, i.e. it must work at full design luminosity
and must tolerate the presence of an underlying event.

The way out may be to use minijets rather than soft hadrons to define the 
rapidity gap. Beyond the different angular distribution of gluons in signal 
and background events, a second distinction is the momentum scale $Q$ of 
the hard process which governs the radiation. In longitudinal
weak boson scattering the color charges, carried by the incident quarks, 
receive momentum transfers given by the transverse momenta of the final 
state quarks, which typically are in the $Q=30$ to 80~GeV range. 
For the background processes, on the other hand, the color charges receive 
a much larger momentum kick, of the order of the weak boson pair mass or even 
the parton center of mass energy of the event, {\it i.e.} $Q\approx 1$~TeV. 
Extra parton emission is suppressed by a factor 
$f_s=\alpha_s {\rm ln}\; (Q^2/p_{T,{\rm min}}^2)$, where $p_{T,{\rm min}}$ is 
the minimal transverse momentum required for a parton to qualify as a jet. 
The jet transverse momentum scale below which multiple minijet emission must 
be expected is set by $f_s={\cal O}(1)$, and this scale is in the 30--50~GeV 
range for the backgrounds but well below 10~GeV for the signal.

For a more precise determination of the typical minijet $p_T$ scale we must 
carry signal and background calculations still one order (in $\alpha_s$) 
higher, i.e. the emission of one additional parton must be included.
Thus we are led to consider $qq\to qqWWg$ and crossing related 
processes\cite{DZ}  for the weak boson scattering signal, and for the 
backgrounds $t\bar tjj$ events\cite{stange} and $WWjj$ events\cite{BHOZ} must 
be simulated. Again the methods discussed in Section~\ref{sec:MC} are 
very useful in performing these tasks. The results of these calculations 
confirm the qualitative arguments made above. For the hard lepton and 
tagging-jet cuts of Eqs.~(\ref{cut1a})--(\ref{cut3}), the $WWjj$ cross 
section, for example, is equal to the lowest order $WWj$ cross section for 
a minimal parton transverse momentum of 37~GeV. Thus multiple
minijets in the 20--50~GeV range are expected for the $WW$ QCD background.
Similar results are found for the $t\bar t$  background while this ``saturation
scale'' for the signal is well below 10~GeV.

For a quantitative estimate of the minijet emission probability, we must
use the higher order programs (which include 
emission of soft partons) in regions of phase space where the $n+1$ jet cross 
section saturates the rate for the hard process with $n$ jets. As the $p_T$ 
of the softest jet is lowered to values where $\sigma (n+1\; {\rm jet}) 
\simeq \sigma (n\; {\rm jet})$, fixed order perturbation theory breaks down and
multiple soft gluon emission (with resummation of collinear singularities into 
quark and gluon structure functions, etc.) needs to be considered in a full
treatment. For the complex processes considered here such calculations are not
yet possible. Instead one can employ the so called 
``truncated shower approximation'' (TSA) to normalize the higher
order emission calculations~\cite{pps}. The tree-level $n+1\,$jet differential 
cross section $d\sigma(n+1\; j)_{\rm TL}$ is replaced by
\begin{equation}\label{tsa}
d\sigma(n+1\; j)_{\rm TSA}=d\sigma(n+1\; j)_{\rm TL}
\left(1-e^{-p_{Tj,min}^2/p_{TSA}^2}\right)\;, \label{reg}
\end{equation}
with the parameter $p_{TSA}$ properly chosen to correctly reproduce the 
lower order $n$~jet cross section when integrated over a given phase space 
region of this hard process. Here $p_{Tj,min}$ is the smallest transverse 
momentum of any of the final state massless partons. As 
$p_{Tj,min}\rightarrow 0$ the final factor in Eq.~(\ref{reg}) acts as a 
regulator, which allows to integrate the $n+1$~parton cross section over the 
full phase space region and simulate $n$-jet and $n+1$-jet events 
simultaneously.

\setlength{\unitlength}{0.7mm}
\begin{figure}[t]
\begin{center}
\input rotate
\vspace*{-0.5in}
\hspace*{0.0in}
\setbox1\vbox{\epsfysize=6.5in\epsffile{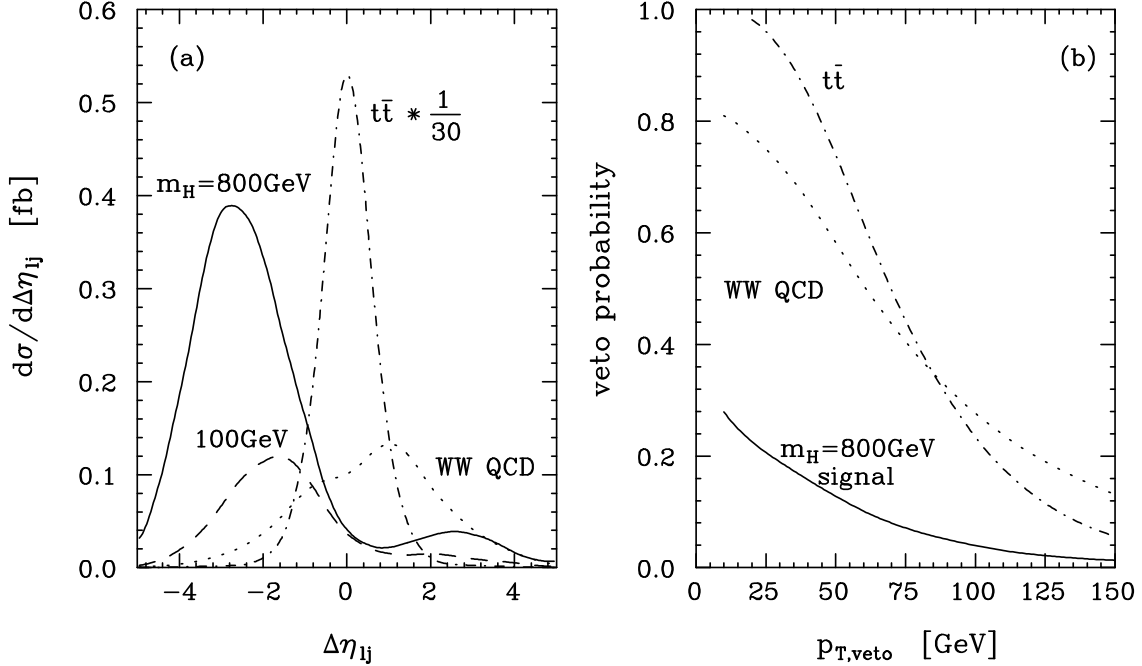}}
\rotl1
\vspace*{-1.2in}
\caption{
Rapidity and transverse momentum distributions of secondary jets.
In (a) $\Delta\eta_{\ell j}$ measures the pseudorapidity distance of the jet 
closest to the leptons from the average lepton rapidity $\bar\eta$. 
Negative values of $\Delta\eta_{\ell j}$ correspond to soft jets on the 
opposite side of the leptons with respect to the tagging jet. The dashed
line shows the distribution for the electroweak background as defined by the 
$m_H=100$~GeV case. The $t\bar tjj$ background has been scaled
down by a factor $30$. The probability to find a veto jet candidate
above a transverse momentum $p_{T,\rm veto}$, in the veto region of 
Eq.~(\protect\ref{cutveto}), is shown in (b). 
{}From Ref.~18. 
\label{fig:veto}
}
\vspace*{-0.1in}
\end{center}
\end{figure}

The results of such a calculation\cite{bpz} are depicted in 
Fig.~\ref{fig:veto}.  The angular distribution of the jet (parton with 
$p_T> 20$~GeV) which is closest to the 
leptons (more precisely, closest to the average lepton rapidity $\bar\eta =
(\eta_{\ell^+}+\eta_{\ell^-})/2$) is shown in Fig.~\ref{fig:veto}(a). 
The background processes favor emission close to the leptons. For the 
Higgs signal this closest jet is typically the second quark in the 
$qq\to qqH$ process and not soft gluon radiation. It is this different 
angular distribution which is at the heart of the rapidity gap trigger. 

A good compromise between strong background rejection and high signal 
acceptance is achieved by vetoing jets in the veto 
region defined by
\begin{equation}\label{cutveto}
p_{Tj}^{\rm veto} > p_{T,\rm veto}\;, \qquad 
\qquad \eta_j^{\rm veto} \varepsilon \;\;
[\eta_\ell^{\rm min}-1.7,\eta_j^{\rm tag}]\;\; {\rm or} \;\;
[\eta_j^{\rm tag},\eta_\ell^{\rm max}+1.7]\; .
\end{equation}
The veto probability as a function of the cut value $p_{T,\rm veto}$ is shown 
in Fig.~\ref{fig:veto}(b). Even though these results were obtained 
in the truncated shower approximation and a more precise modeling is needed,
they clearly demonstrate that, in the central region, the backgrounds have a 
much higher probability to produce additional minijets from QCD radiation
than the weak boson scattering signal.
Even for the $t\bar t$ background, where a strong suppression is obtained by
vetoing the central $b$-quark jets arising from the top decays, the veto on 
the jet activity from soft QCD radiation provides an additional 
suppression by a factor 2 for $p_{T,\rm veto}=20$~GeV. Final cross 
section values for signal and backgrounds are given in the last column of  
Table~\ref{tab:Hcross}.  Should minijet vetoing be possible at the LHC for 
even smaller $p_{T,\rm veto}$ values then
the $t\bar t$ background to $H\to WW$ events can effectively be eliminated.

\section{Conclusions}

Rapidity gaps in hard dijet events at the Tevatron and minijet patterns in 
weak boson scattering at the LHC have their common origin in processes which
are dominated by $t$-channel color singlet exchange. Both can be understood 
in terms of soft gluon radiation patterns which are determined by the color 
structure of the underlying hard scattering event. At the lower energy of 
the Tevatron the soft gluon radiation pattern is reflected in the distribution
of soft hadrons in events without a residual minimum bias event (underlying
event). Going up one order of magnitude to LHC energy, ``soft gluons'' can
be hard enough to be seen as distinct minijets, and this opens new  
strategies for the study  of weak boson scattering events.

The theoretical description of these multi-parton processes has become
possible with the advent of amplitude techniques. By directly evaluating
polarization amplitudes numerically, fast numerical programs can be designed 
which remain efficient for processes with large numbers of Feynman graphs.
In addition, code generation for SM processes has now been automated and this
allows us to concentrate on the physics issues in the analysis of multi-parton
processes.

Via the study of rapidity gap events we are obtaining important information
for future LHC physics studies already now at the Tevatron. Similarly, 
multiple 
minijet emission, which is expected to be quite common at the LHC, can already
now be studied in the hardest ``dijet'' events at the 
Tevatron~\cite{CDFmultij,SZ}. Both examples
show the important interplay between present experiments and new theoretical 
ideas and tools in multi-parton analysis. Some of these developments were 
motivated by the search for new physics at future colliders. At the same time
they aid in the preparation of experiments at these machines.

\section*{Acknowledgements}
Special thanks go to the organizers for making this most enjoyable and 
stimulating meeting possible. In preparing this talk I have benefitted 
from many discussions with V.~Barger, J.~D.~Bjorken, A.~Brandt, F.~Halzen, 
W.~Long, R.~Phillips, and D.~Summers and I want to thank them for sharing 
their insights with me. I would like to thank T.~Stelzer for teaching 
me about MadGraph and for a critical reading of Section~3. 
This research was supported in part by the University of Wisconsin Research
Committee with funds granted by the Wisconsin Alumni Research Foundation,
and by the U.~S.~Department of Energy under Grant No.~DE-FG02-95ER40896.

\newpage

\end{document}